\documentclass[11pt, oneside]{scrartcl}   	
\usepackage{geometry}                		
\usepackage{latexsym}
\usepackage{graphicx}
\usepackage{amsmath}
\usepackage{amssymb}
\usepackage{slashed}
\usepackage{color}
\usepackage[dvipsnames, table]{xcolor}
\usepackage{mathtools}
\usepackage{siunitx}
\usepackage{xspace}
\usepackage[colorlinks=true,a4paper=true,backref=false,linktocpage=true,citecolor=blue,urlcolor=blue,linkcolor=blue,pdfpagemode=UseOutlines]{hyperref}
\newcommand{\gcl}{\cellcolor[gray]{0.90}}
\usepackage{multirow}
\usepackage{comment}
\usepackage{booktabs}

\newcommand{\Section}[1]{\section[\boldmath #1]{\boldmath #1}}

\newcommand{\Subsection}[1]{\subsection[#1]{\boldmath #1}}

\newcommand{\be}{\begin{equation}}
\newcommand{\ee}{\end{equation}}
\newcommand{\bea}{\begin{eqnarray}}
\newcommand{\eea}{\end{eqnarray}}

\newcommand{\mc}{\mathcal}

\newcommand{\nn}{\nonumber}

\newcommand{\Ga}{\Gamma}

\newcommand{\De}{\Delta}

\newcommand{\Bs}{\ensuremath{B^0_s}\xspace}

\newcommand{\re}{{\rm Re}}
\newcommand{\im}{{\rm Im}}
\newcommand{\ADG}{A_{\De\Ga}}

\newcommand{\GeV}{{\rm GeV}}

\newcommand{\Bsmmy}{\ensuremath{B_s \to \mu^+ \mu^- \gamma}\xspace}
\newcommand{\mm}{\ensuremath{\mu^+\mu^-}\xspace}
\newcommand{\Bsmm}{\ensuremath{B_s \to \mm}\xspace}

\newcommand{\mumu}{\mu^{+} \mu^{-}}

\newcommand{\Bstog}{\ensuremath{B_s \to \gamma}\xspace}

\newcommand{\uemu}{u(e,\mu)}

\newcommand{\FA}{\ensuremath{V_\parallel}\xspace}

\newcommand{\FV}{\ensuremath{V_\perp}\xspace}

\newcommand{\FTA}{\ensuremath{T_\parallel}\xspace}
\newcommand{\FTV}{\ensuremath{T_\perp}\xspace}

\newcommand{\FTVA}{\ensuremath{T_{\perp,\parallel}}}

\def\btosmm{\ensuremath{b \to s \mu^+ \mu^-}\xspace}
\def\btosg{\ensuremath{b \to s \gamma}\xspace}

\newcommand{\RKKst}{\ensuremath{R_{K^{(*)}}}\xspace}
\newcommand{\RDDst}{\ensuremath{R{(D^{(*)})}}\xspace}
\newcommand{\Cten}{\ensuremath{C_{10}}\xspace}
\newcommand{\Cnine}{\ensuremath{C_{9}}\xspace}

\newcommand{\Bz}{\ensuremath{B^0}\xspace}
\newcommand{\Kstarz}{\ensuremath{K^{*0}}\xspace}
\newcommand{\Kstar}{\ensuremath{K^{*}}\xspace}

\DeclareOldFontCommand{\rm}{\normalfont\rmfamily}{\mathrm}
\DeclareOldFontCommand{\sf}{\normalfont\sffamily}{\mathsf}
\DeclareOldFontCommand{\tt}{\normalfont\ttfamily}{\mathtt}
\DeclareOldFontCommand{\bf}{\normalfont\bfseries}{\mathbf}
\DeclareOldFontCommand{\it}{\normalfont\itshape}{\mathit}
\DeclareOldFontCommand{\sl}{\normalfont\slshape}{\@nomath\sl}
\DeclareOldFontCommand{\sc}{\normalfont\scshape}{\@nomath\sc}

\definecolor{dartmouthgreen}{rgb}{0.05, 0.5, 0.06}

\newcommand{\reffig}[1]{Fig.\,\ref{#1}}

\newcommand{\reftab}[1]{Table\,\ref{#1}}
\newcommand{\refsec}[1]{Sec.\,\ref{#1}}

\newcommand{\invfb}{\ensuremath{~\mathrm{fb}^{-1}}\xspace}

\begin{document}

\begin{flushright}
\small
LAPTH-039/23
\end{flushright}
\vskip0.5cm

\begin{center}
{\sffamily \bfseries \LARGE \boldmath
Insights on the current semi-leptonic $B$-decay\\
[0.2cm]discrepancies -- and how $B_s \to \mu^+ \mu^- \gamma$ can help
}\\[0.8 cm]
{\normalsize \sffamily \bfseries Diego Guadagnoli$^1$, Camille Normand$^1$, Silvano Simula$^2$, Ludovico Vittorio$^1$} \\[0.5 cm]
\small
$^1${\em LAPTh, Universit\'{e} Savoie Mont-Blanc et CNRS, Annecy, France}\\
[0.1cm]
$^2${\em INFN, Sezione di Roma Tre, Via della Vasca Navale 84, 00146 Rome, Italy}
\end{center}

\medskip

\begin{abstract}

\noindent $B_s \to \mu^+ \mu^- \gamma$, measured at high $q^2$ as a partially reconstructed decay, can probe the origin of the existing discrepancies in semi-leptonic $b \to s$ and $b \to c$ decays. We perform a complete study of this possibility. We start by reassessing the alleged discrepancies, with a focus on a unified EFT description. Using the SMEFT, we find that the tauonic Wilson coefficient required by $R(D^{(*)})$ implies a universal muonic Wilson coefficient of precisely the size required by semi-muonic BR data and, separately, by semi-muonic angular analyses. We thus identify reference scenarios. Importantly, $B_s \to \mu^+ \mu^- \gamma$ offers a strategy to access them without being affected by the long-distance issues that hamper the prediction of semi-leptonic $B$ decays at low $q^2$. After quantifying to the best of our knowledge the $B_s \to \mu^+ \mu^- \gamma$ experimental over the long haul, we infer the $B_s \to \mu^+ \mu^- \gamma$ sensitivity to the couplings relevant to the anomalies. In the example of the real-$\delta C_{9,10}$ scenario, we find significances below 3$\sigma$. Such figure is to be compared with other {\em single}-observable sensitivities that one can expect from e.g. BR and angular data, whether at low or high $q^2$, and not affected by long-distance issues such as narrow resonances or intermediate charmed di-meson rescattering.

\end{abstract}

\date{} 

\newpage

\tableofcontents

\Section{Introduction} \label{sec:intro}

An ensemble of $b \to s \mm$ data including branching-ratio measurements as well as angular ones displays a less than perfect agreement with the SM prediction. Within LHCb, the concerned observables include $\mc B(B \to M \mm)$ for $M = K^{0,+,*+}$~\cite{LHCb:2014cxe}, $K^{*}(892)$~\cite{LHCb:2016ykl}, $K^{*0}$~\cite{LHCb:2020lmf}; $\mc B(B^0_s \to \phi \mm)$~\cite{LHCb:2015wdu,LHCb:2021zwz}; the baryonic channels $\mc B(\Lambda_b \to \{\Lambda, \Lambda(1520)\} \mm)$~\cite{LHCb:2015tgy,LHCb:2018jna,LHCb:2023ptw}; two angular analyses for $M = K^{*+, *0}$~\cite{LHCb:2020gog,LHCb:2020lmf}. The trend suggested by these datasets is generally, although not uniformly, confirmed by measurements collected by Atlas, BaBar, Belle, CMS~\cite{Belle:2009zue,BaBar:2012mrf,CMS:2013mkz,Belle:2016fev,CMS:2017rzx,ATLAS:2018gqc,CMS:2020oqb}. Finally, past experimental indications of lepton-universality violation have disappeared~\cite{LHCb:2022qnv}. Hence the above tensions, if confirmed, are not expected to concern dimuons channels only.

Importantly, this disagreement is seen only for low di-muon invariant mass squared $q^2$, and is consistent across the whole ensemble, in the sense that the same Wilson-coefficient shift accommodates all discrepancies. However, interpreting this shift as due to new physics (NP) relies on a strong assumption about the possible size of long-distance contributions to the different observables concerned. These contributions are, on the one side, difficult to estimate, and on the other side largely {\em equivalent} to (i.e. parametrically interchangeable with) the NP shifts that one wants to probe. 

A complete estimation of such long-distance effects is, as well-known, an issue as important as it is challenging. The existing calculations in Refs.\,\cite{Gubernari:2020eft,Gubernari:2022hxn} (building on Refs.~\cite{Khodjamirian:2010vf,Khodjamirian:2012rm,Bobeth:2017vxj}) focus on the ``charm-loop''-to-$\gamma^*(q^2)$ amplitude, whose long-distance effects correspond to poles and cuts in the $q^2$ variable. On the other hand, Ref.~\cite{Ciuchini:2022wbq} (see also Refs.~\cite{Ciuchini:2015qxb, Ciuchini:2017mik, Ciuchini:2019usw, Ciuchini:2020gvn, Ciuchini:2021smi}) emphasizes the importance of including contributions from $B$ to di-meson rescatterings, that correspond to cuts in the full decay variable $(q+k)^2$, where $k$ is the momentum of the final-state $K^{(*)}$.\footnote{The only existing phenomenological estimate \cite{Ladisa:2022vmh}, considering light di-meson intermediate states (i.e. not $D \bar D^{*}$ ones), suggests that such contributions may be sizeable.} Ideally, one should start from an amplitude that is function of $q^2$ and of $(q+k)^2$, such that cuts in the second variable reproduce $B \to D \bar D^{*}$ decay rates, and only then set $(q+k)^2 = m_B^2$. In other words, inclusion of $(q+k)^2$ as a variable should allow to take into account the so-called \emph{anomalous cut}\footnote{See e.g. discussion in Ref.\,\cite{Lucha:2006vc}.} in addition to the ``usual'' cut associated to the $c \bar c$ threshold. The state-of-the-art calculations in Refs.~\cite{Gubernari:2020eft,Gubernari:2022hxn} allow for complex-valued helicity amplitudes (see e.g. discussion in Ref.~\cite{Bobeth:2017vxj}), which are nevertheless functions of $q^2$ only, and are denoted as $\mc H_\lambda(q^2)$. We are lacking a proof that the theoretical and experimental input used to constrain complex $\mc H_\lambda(q^2)$ fully encapsulates the structure due to the cuts in $(q+k)^2$. It is clear that going beyond the calculations in Refs.~\cite{Gubernari:2020eft,Gubernari:2022hxn}---that constitute a benchmark for all that is calculable with regards to this issue---is a daunting task, primarily because there is no known EFT that would allow a quantitative estimate of the anomalous-cut contributions.

There are two clear avenues towards resolving the above quandary: on the theory side, an estimate of the mentioned missing long-distance effects, dispelling any doubt that they may be mimicking NP;
on the experimental side, the consideration of alternative observables---in particular, in alternative $q^2$ regions---that are sensitive to the very same short-distance physics without being affected by the long-distance effects in question.

The branching ratio for \Bsmmy, measured at high $q^2$ as a partially reconstructed decay \cite{Dettori:2016zff} (see Ref. \cite{LHCb:2021vsc,LHCb:2021awg} for a first application at LHCb)  offers an example of such an alternative observable. It actually provides a litmus test between the two mentioned explanations: NP versus long-distance effects mimicking it. In fact, at high $q^2$ this observable's long-distance contributions are dominated by two form factors that should be accessible by first-principle methods in the short term---whereas the helicity amplitudes that are input of the presently discrepant $b \to s \mm$ data do not enter at all.

A clear test then emerges: one may predict $\mc B(\Bsmmy)$ assuming NP shifts as large as suggested by the currently discrepant low-$q^2$ $b \to s \mumu$ observables if the incalculable long-distance contributions to these observables are negligible. A $\mc B(\Bsmmy)$ measurement that confirms our prediction would  circumstantially support NP shifts of that size. This paper aims at discussing the relation between NP sensitivity and amount of data required for such test. To this end, the paper is structured as follows. 
In \refsec{sec:global-fits}, we re-assess the $b\to s \mm$ discrepancies by performing global likelihood fits of all relevant observables and their up-to-date measurements, among the others \RKKst and \Bsmm. From such fits we extract NP shifts to the semi-leptonic Wilson coefficients known as $\Cnine^{bs\mu\mu}$ and $\Cten^{bs\mu\mu}$, to be used as reference for the rest. We consider separately the cases of real and complex shifts to these Wilson coefficients, and we examine to what extent data (including $b \to c \ell^- \bar \nu$) obey a coherent effective-theory picture. In \refsec{sec:Bsmmy-exp-SM-NP}, we attempt an extrapolation of the experimental and theoretical uncertainties associated to \Bsmmy at high $q^2$ and we use such extrapolation to project the NP sensitivity of the observable, using as NP benchmarks the couplings discussed in \refsec{sec:global-fits}. We draw our conclusions in \refsec{sec:concl}. Finally, the \hyperref[app:SMEFT]{Appendix} collects additional information and plots related to the global fits in \refsec{sec:global-fits}.

\Section{NP benchmarks} \label{sec:global-fits}

\Subsection{Real Wilson-coefficient shifts}

\begin{table}[h!]
  \renewcommand{\arraystretch}{1.2}
  \centering
  \begin{tabular}{|l|c|c|}
    \hline
    &  \bf Refs. measurements                          &  \bf Refs. prediction     \\
    \hline
    \gcl $ \pmb{b \to s \mu^+ \mu^-}$ \textbf{BR obs.\phantom{xxxxx}}           & \gcl   & \gcl  \\
    $\left< \frac{d \mc B}{dq^2}\right> (B^+\to K^{(*)}\mu\mu )$    & \cite{LHCb:2014cxe,CDF:2012qwd}  &  \cite{Bharucha:2015bzk,Gubernari:2018wyi,Parrott:2022rgu} \\
    $\left<\frac{d\mc B}{dq^2}\right>(B_0\to K\mu\mu)$          &   \cite{CDF:2012qwd,CMS:2015bcy,LHCb:2014cxe}      &  \cite{Parrott:2022rgu}    \\
    $\left< \frac{d \mc B}{dq^2} \right> (B_s\to \phi\mu\mu)$ & \cite{LHCb:2015wdu,CDF:2012qwd,LHCb:2021zwz} & \multirow{2}{*}{\cite{Bharucha:2015bzk}}\\
    $\left< \frac{d \mc B}{dq^2} \right> (B_0\to K^*\mu\mu)$ & \cite{CMS:2015bcy,LHCb:2016ykl}         &     \\
    $\left<\mc B\right>(B\to X_s\mu\mu)$                           & \cite{BaBar:2013qry}           &  \cite{Huber:2015sra}   \\
    \hline
    \gcl  $ \pmb{b \to s \mu^+ \mu^-}$ \textbf{angular and CPV obs.\phantom{xxxxx}}           & \gcl   & \gcl   \\
    $\left<F_L,P_{1},P_{4,5}^{\prime},A_{{\rm FB}}\right>(B_0\to K^* \mu^+ \mu^-)$     & \cite{CDF:2012qwd,ATLAS:2018gqc,LHCb:2020lmf,CMS:2017ivg}   &  \multirow{4}{*}{\cite{Bharucha:2015bzk}}   \\
    $\left< F_L,P_{1,2},P'_{4,5}\right>(B^+\to K^{*+} \mu^+ \mu^-)$   & \cite{LHCb:2020gog}                             &   \\
    $\left< F_L,S_{3,4,7}\right>(B_s\to \phi\mu\mu)$ &  \cite{LHCb:2021xxq} &  \\
    $A_{3-9}(B_0 \to K^* \mu^+ \mu^-)$ & \cite{LHCb:2015svh} &   \\
    \hline                                                                                                                                   
    $\pmb{R_{K/K^*}}$     & \cite{Belle:2019oag,BELLE:2019xld,LHCb:2022qnv,LHCb:2022zom} &   \cite{Bharucha:2015bzk,Gubernari:2018wyi,Parrott:2022rgu}    \\
    \hline
    $\pmb{\mc B(B_{d,s}\to \mu\mu)}$    & \cite{ATLAS:2018cur,LHCb:2021vsc,LHCb:2021awg,CMS:2022mgd} &  \cite{DeBruyn:2012wj}     \\
    \hline   
    \gcl  $ \pmb{b \to s \gamma}$ \textbf{obs.\phantom{xxxxx}}           & \gcl   & \gcl   \\
    $\left<\mc B, A_{CP}\right>(B\to X_s\gamma)$     &     \cite{Misiak:2017bgg} &   \cite{Misiak:2006ab,PhysRevLett.114.221801}        \\
    $\mc B(\Bz\to\Kstarz\gamma)/\mc B(\Bs\to\phi\gamma)$     &     \cite{Aaij:2012ita} &    \cite{Bharucha:2015bzk,Gubernari:2018wyi}       \\
    $\mc B(B\to\Kstar\gamma)$     &    \cite{Amhis:2014hma} &         \multirow{4}{*}{\cite{Bharucha:2015bzk}}  \\
    $\mc B(\Bs\to\phi\gamma)$     &    \cite{Dutta:2014sxo} &           \\
    $\ADG,S(\Bs\to\phi\gamma)$     &    \cite{Aaij:2019pnd} &           \\
    $S_{\Kstarz\gamma}$     &      \cite{Amhis:2014hma} &           \\
    \hline
  \end{tabular}
  \caption{List of the most constraining observables and their measurements implemented in the {\tt flavio v2.5.4} Python package at the date of publication. For the exclusive channels, the predictions refer to the particular set of form factors of the relevant transition. Helicity amplitudes are detailed in Refs.~\cite{Descotes-Genon:2013zva,Jager:2012uw,PhysRevD.93.054008}.}
  \label{tab:observables-list}
  \renewcommand{\arraystretch}{1.0} 
\end{table}
One can consider the list of measurements\footnote{
The $\Lambda_b \to \Lambda \mu^+ \mu^-$ as well as $\Lambda_b \to \Lambda(1520) \mu^+ \mu^-$ measurements in Refs.~\cite{LHCb:2015tgy, LHCb:2018jna} and in Ref.~\cite{LHCb:2023ptw}, respectively, may be tested against the pioneering SM calculations in Ref.~\cite{Detmold:2016pkz} and in Refs.~\cite{Meinel:2020owd, Meinel:2021mdj}, respectively. In this work, however, we do not include $b \to s$ data from semileptonic decays of baryons following the guidelines of the latest FLAG Review \cite{FlavourLatticeAveragingGroupFLAG:2021npn}.} in \reftab{tab:observables-list} and perform a global fit, using the python software package \texttt{flavio}\footnote{We use version \texttt{2.5.4}.} \cite{Straub:2018kue}. As a first step, we focus on shifts to semi-leptonic Wilson coefficients involving dimuons, defining $(k=\,9,\,10)$
\bea
\label{eq:def-NP-shifts}
\begin{tabular}{llllll}
{\footnotesize NP shift} & & {\footnotesize $\ell$-specific + $\ell$-univ. parts} & & & {\footnotesize full WC (NP + SM)} \\
\hline\\[-0.3cm]
$\delta C_k^{bsee}$ 	    & $\equiv$ & $\delta C_k^{(e)} + \delta C_k^{u(e,\mu)}$ & $=$ & $C_k^{bsee} - C_k^{bs\ell\ell,\rm SM} ~,$ & $C_k^{bsee} \equiv C_k^{(e)}~,$\\
$\delta C_k^{bs\mu\mu}$   & $\equiv$ & $\delta C_k^{(\mu)} + \delta C_k^{u(e,\mu)}$ & $=$ & $C_k^{bs\mu\mu} - C_k^{bs\ell\ell,\rm SM} ~,$ & $C_k^{bs\mu\mu} \equiv C_k^{(\mu)}~,$\\
$\delta C_k^{bs\tau\tau}$ & $\equiv$& $\delta C_k^{(\tau)}$ & $=$ & $C_k^{bs\tau\tau} - C_k^{bs\ell\ell,\rm SM}~,$ & $C_k^{bs\tau\tau} \equiv C_k^{(\tau)}~,$
\end{tabular}
\eea
with the SM part also abbreviated through $C_k^{bs\ell\ell,\rm SM} \equiv C_k^{(\ell), \rm SM}$. As highlighted in eq.~(\ref{eq:def-NP-shifts}), NP shifts are identified from a leading $\delta$ (at variance with much of the literature). In addition, the full NP shift generally consists of a lepton-specific component, labeled as $^{(e), (\mu), (\tau)}$ plus a lepton-universal one. In most of our discussion we will actually focus on universal components only concerning the light leptons, i.e. on {\em light-lepton-}universal shifts, that are labeled with~$^{\uemu}$.

\begin{figure}[th!]
  \begin{center}
  \includegraphics[width=0.45\linewidth]{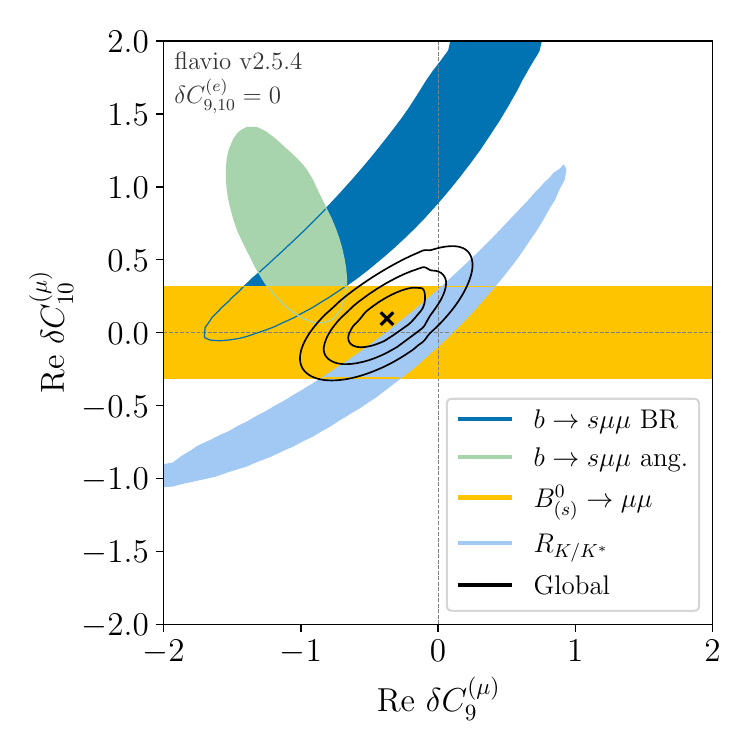}
  \includegraphics[width=0.45\linewidth]{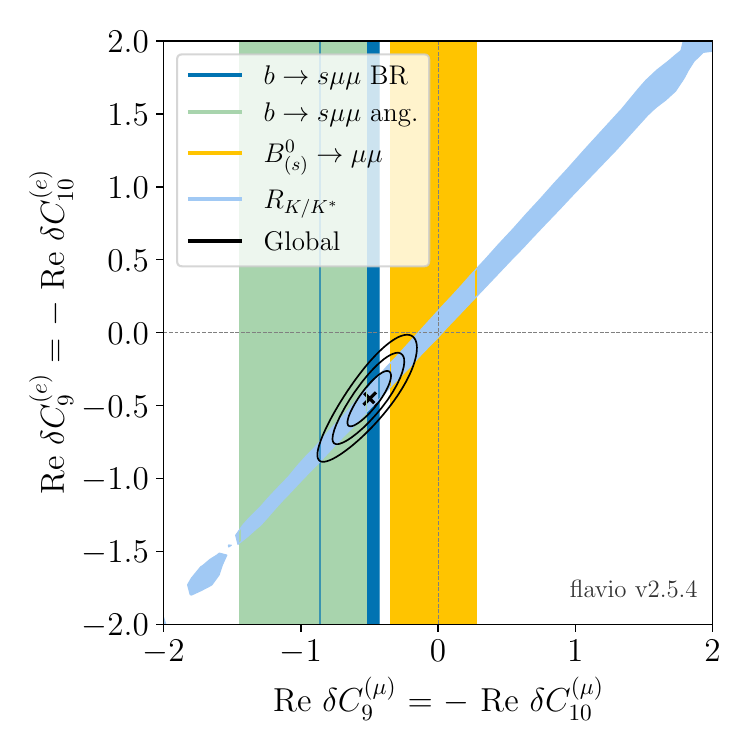}
  \includegraphics[width=0.45\linewidth]{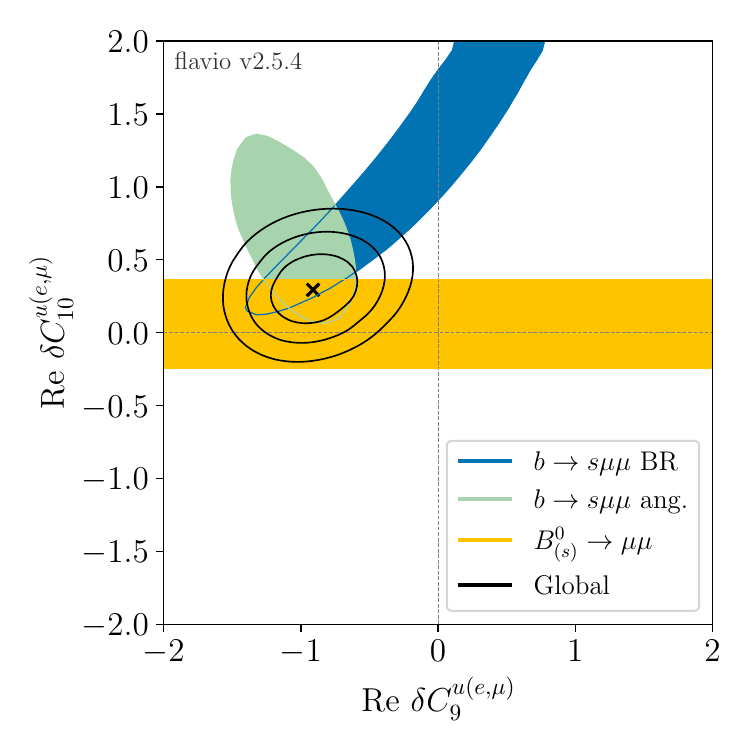}
  \includegraphics[width=0.45\linewidth]{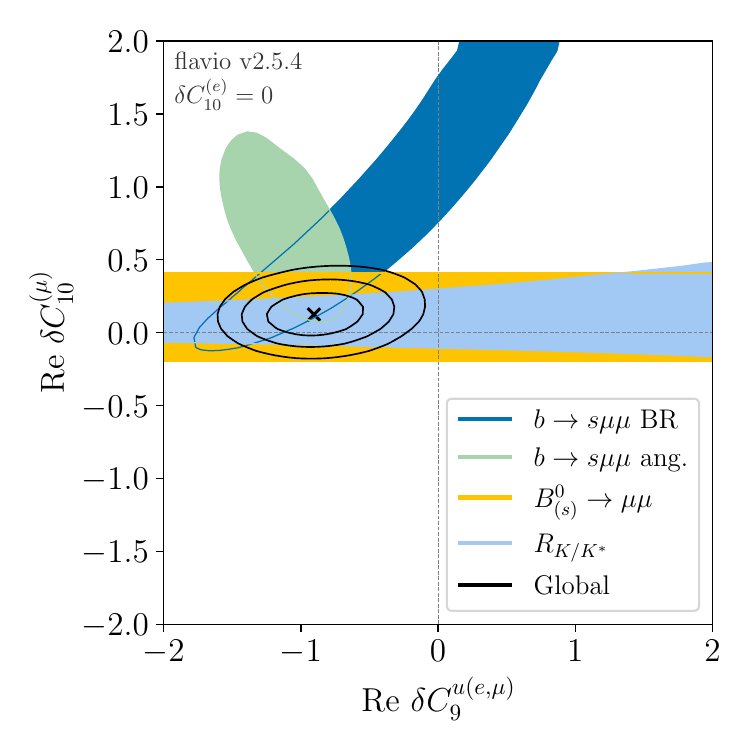}
  \caption{Global fits to the Wilson coefficients 
  $\delta C_9^{(\mu)}$ vs. $\delta C_{10}^{(\mu)}$, with the assumption $\delta C_{9,10}^{(e)} = 0$ (top left); 
  to $\delta \Cnine^{(\mu)} = - \delta \Cten^{(\mu)}$ vs. $\delta \Cnine^{(e)} = - \delta \Cten^{(e)}$ (top right);
  to $\delta C_{9}^{\uemu}$ vs. $\delta C_{10}^{\uemu}$ (bottom left);
  to $\delta \Cnine^{\uemu}$ vs. $\delta \Cten^{(\mu)}$ (bottom right).}
  \label{fig:fit-9vs10}
  \end{center}
\end{figure}

With eq.~(\ref{eq:def-NP-shifts}) setting the notation, we show the case $\delta C_9^{(\mu)}$ vs. $\delta C_{10}^{(\mu)}$ in the top-left panel of \reffig{fig:fit-9vs10}. The fit suggests that the updated \RKKst measurement and the anomalies observed in the $b \to s \mm$ sector---branching ratios (BRs) and angular observables---do not yield a coherent picture in a scenario where one only shifts $C_9^{(\mu)}$ and $C_{10}^{(\mu)}$, i.e. where their counterparts for all other flavours are assumed SM-like.\footnote{\label{foot:WCs}For conciseness, and to align with conventions in the literature, the Wilson coefficient shown in plots is the beyond-SM shift. For reference the SM values are $C_9^{(\ell), {\rm SM}} = 4.327$, $C_{10}^{(\ell), {\rm SM}} = -4.262$ (see e.g. Ref.~\cite{Beneke:2020fot} for accurate notation on the scale and the short-distance contributions included). Also, $C_7^{\rm SM} = -0.303$, to be used around eq.~(\ref{eq:C7_def}).}
Specifically, the combination of the recent \Bsmm measurements by both LHCb and CMS collaborations provides a strong constraint on $\delta C_{10}^{(\mu)}$, which is now consistent with zero. Besides, the \RKKst measurements on the one side and the discrepant $b \to s \mm$ observables on the other side, each constrain the $\delta C_{9}^{(\mu)}$ vs. $\delta C_{10}^{(\mu)}$ plane obliquely, but the two respective regions overlap at no better than 2$\sigma$. These findings may be contrasted with the analysis in Refs.~\cite{Greljo:2022jac,Altmannshofer:2023uci}.

As a second case of interest, we consider a scenario that is especially appealing in the light of a UV interpretation, where NP fulfils the constraint $\delta \Cnine^{(\ell)} = - \delta \Cten^{(\ell)} \equiv \delta C_{LL}^{(\ell)}/2$,\footnote{This identification follows trivially from writing the effective Hamiltonian either in the ``traditional basis'', $\propto C_9 \mc O_9 + C_{10} \mc O_{10}$ with $\mc O_9 \propto (\bar s \gamma^\mu_L b) \, (\bar \ell \gamma_{\mu} \ell)$ and  $\mc O_{10} \propto (\bar s \gamma^\mu_L b) \, (\bar \ell \gamma_{\mu} \gamma^5 \ell)$, or in the chiral basis, $\propto C_{LL} \mc O_{LL}$, where $\mc O_{LL} \propto (\bar s \gamma^\mu_L b) \, (\bar \ell \gamma_{\mu L} \ell)$. Here we omit flavour indices for simplicity, as well as proportionality factors immaterial to the identification.} in both the muonic and electronic channels. This is shown in the top-right panel of \reffig{fig:fit-9vs10}, again displaying that the region favoured by \Bsmm is in mild tension with the regions preferred by $b \to s \mumu$ BR and angular data---that instead are mutually consistent.

The very SM-like \RKKst measurements can be reconciled with the $b \to s \mm$-sector BR and angular observables by allowing NP in the electron channels. Hence a first natural scenario to account for discrepant BR and angular data, and for {\em no} LFUV in $b \to s \ell^+ \ell^-$ ($\ell = e, \mu$) is the case $\delta \Cnine^{\uemu}$ vs. $\delta \Cten^{\uemu}$ (we refer again to eq.~(\ref{eq:def-NP-shifts}) and the surrounding text for the meaning of $\uemu$). This case is displayed in the bottom-left panel of \reffig{fig:fit-9vs10}. In this panel the \RKKst constraint is trivially satisfied in the whole plane and thus not displayed. What we deem significant in this scenario is the fact that BR (in blue) and angular data (in green) constrain this WC plane in two independent directions, and the preferred regions of either set neatly overlap at a non-zero value of $\delta C_9^{u(e,\mu)}$---while staying consistent with a SM-like, i.e. null within errors, value of $\delta C_{10}^{u(e,\mu)}$. As a result, this scenario is consistent with all observables at the $1\sigma$ level.
We note that the hint at $\delta C_{10}^{(\mu)}$ consistent with zero holds irrespective of its relation to $\delta C_{10}^{(e)}$. In the last two panels of \reffig{fig:fit-9vs10} we assume $\delta C_{10}^{(e)} = \delta C_{10}^{(\mu)}$ and, respectively, $\delta C_{10}^{(e)} = 0$. The main difference between these two fits is in the significance of the null solution for the WC on the $y$ axis, slightly above 1$\sigma$ and respectively within 1$\sigma$ in the bottom-left and bottom-right panels. The difference is clearly due to the \RKKst constraint, which as already mentioned is ineffectual in the bottom-left panel.
\label{page:two-messages}

As a whole, \reffig{fig:fit-9vs10} delivers {\em two} clear-cut messages from data:
\begin{itemize}
\item[{\em (a)}] the \Bsmm update disfavors shifts to muonic $\delta C_{10}$ and the \RKKst measurement favors a solution where the muonic vs. electronic WC shifts are equal. Jointly, these pieces of information would tend to disfavor the $\delta C_{LL}^{(\ell)}$ scenario (top-right panel of \reffig{fig:fit-9vs10}). We will see, however, that this is true only if the shift is weak-phase-aligned with the SM---the relevant discussion is in \refsec{sec:complex_WCs};
\item[{\em (b)}] BR \& angular data {\em separately} point towards a non-zero, but {\em light-lepton-}universal shift $\delta C_9^{u(e,\mu)}$.
\end{itemize}

We deem the last finding non-trivial. In the next section we will see that it can be understood within a SMEFT picture, that relates $b \to s$ and $b \to c$ discrepancies and that turns out to quantitatively account for both.

\Subsection{The SMEFT-induced connection between $b \to s$ and $b \to c$ anomalies: an update} \label{sec:SMEFT}

While the $\delta C_9^{u(e,\mu)}$ shift just discussed has to be corroborated by further data (e.g. updates in the channels of Refs.~\cite{LHCb:2014cxe,LHCb:2016ykl}), it leaves a testable imprint already. In fact, quite generally a universal $\delta C_9$\footnote{\label{foot:universal}We do not use $\delta C_9^{\uemu}$ to denote such shift, because the considerations in these paragraphs generally apply to a $\delta C_9$ shift common to all three leptonic generations. As a rule, when ``universal'' is used textually rather than as $^{\uemu}$, we mean that it applies to all three generations.} tends to correlate with effects in {\em charged-current} semi-leptonic transitions, in particular $b \to c \ell^- \bar \nu$. This mechanism is to be expected as consequence of the inescapable RG running between the scale of the new dynamics and the $b$ scale, to the extent that the new dynamics is sizeably above the EW symmetry-breaking threshold.

As a matter of fact, the possibility of a low-scale universal $C_9$ shift arising from high-scale interactions has been pointed out within the WET \cite{Bobeth:2011st}, the SMEFT \cite{Aebischer:2019mlg} and in the context of renormalizable UV models \cite{Crivellin:2018yvo}. A particularly plausible example is that of semi-tauonic, $SU(2)_L$-symmetric SMEFT operators, that will leave a direct imprint on $R(D^{(*)})$. The corresponding SMEFT WCs are denoted as $[C_{lq}^{(1), (3)}]_{ijmn}$\footnote{We adopt the ``Warsaw basis'' \cite{Grzadkowski:2010es} and the convention that $\ell$ and $l$ refer respectively to charged leptons below the EW scale and to the lepton doublets above the EW scale, see e.g. \cite{Aebischer:2017ugx,Aebischer:2019mlg}.} with $ij$ leptonic, $mn$ quark indices, and $^{(1),(3)}$ labeling the $SU(2)_L$-singlet or triplet instances. The latter are customarily assumed to be equal in order to automatically fulfil $B \to K^{(*)} \nu \bar \nu$ constraints~\cite{Buras:2014fpa}. We adhere to such assumption and drop the $^{(1),(3)}$ index, so that the SMEFT WCs will be simply denoted as $C_{ii23}$. It is then interesting to study the implications that different assumptions on $ii=11$ vs. $22$ vs. $33$ have on the global consistency between $b \to s$ and $b \to c$ discrepancies. For example, $C_{ii23}$ generates a matching contribution to $\delta C_{9}^{(\ell_i)} = -\delta C_{10}^{(\ell_i)}$; in turn $C_{3323}$ can {\em also} contribute a sizeable lepton-universal $\delta C_9$ shift. (In principle, every $ii$ contributes such RG-induced lepton-universal $\delta C_9$ shift. However, $ii=33$ is the least constrained in the light of data and can thus afford to provide the dominant contribution.)

\begin{figure}[th!]
  \begin{center}
  \includegraphics[width=0.45\linewidth]{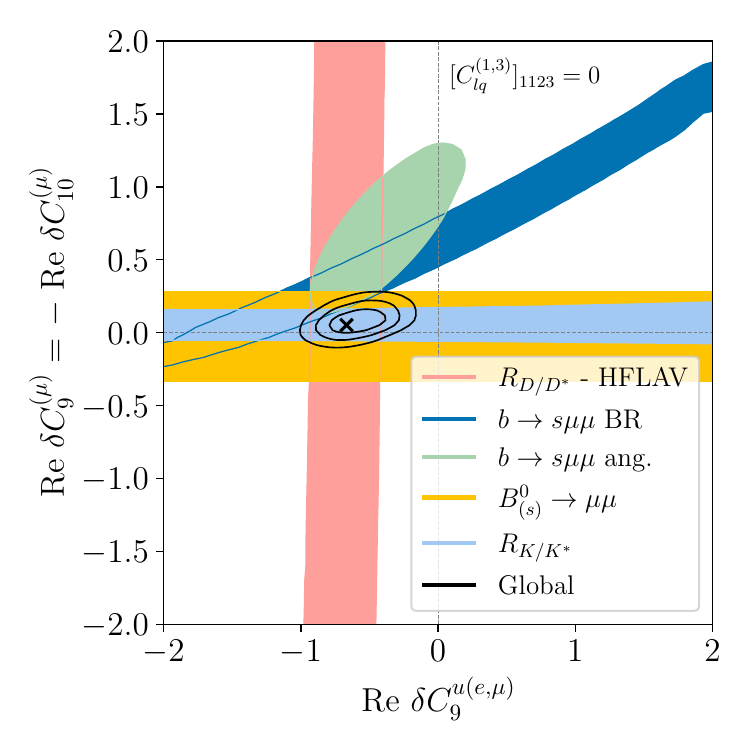}
  \includegraphics[width=0.45\linewidth]{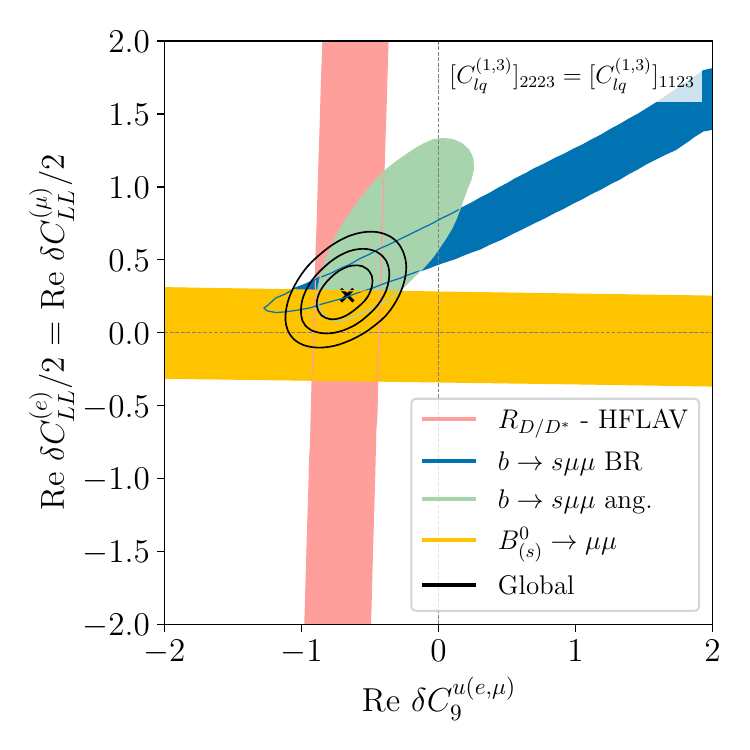}
  \includegraphics[width=0.45\linewidth]{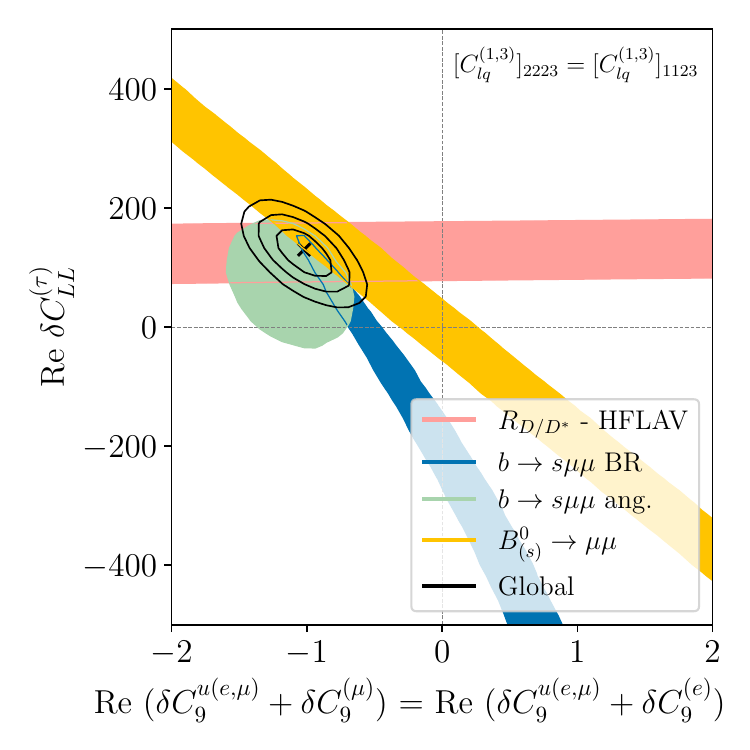}
  \includegraphics[width=0.45\linewidth]{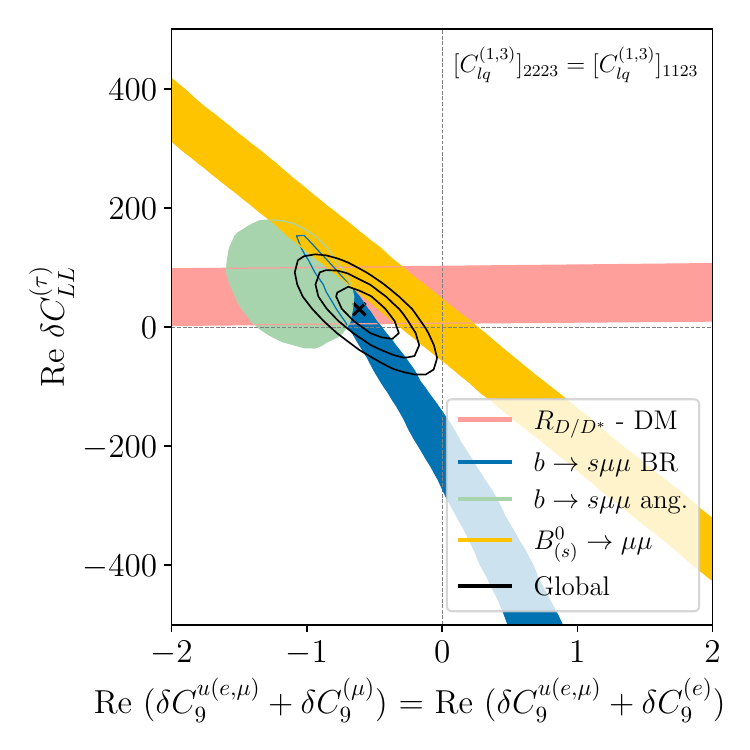}
  \caption{Semileptonic $b \to s$ {\em and} $b \to c$ constraints in the plane $\delta C_9^{\uemu}$ vs. $\delta C_{LL}^{(\mu)}$ obtained from SMEFT WCs under the assumptions $|C_{3323}| \gg |C_{2223}| > 0$, and $C_{1123} = 0$ (top-left panel) or $|C_{3323}| \gg |C_{2223} = C_{1123}| > 0$ (top-right panel). The latter scenario is also shown in the plane of total $\delta C_9$, equal for {\em light} leptons, vs. $\delta C_{LL}^{(\tau)}$ (bottom-left panel) and with the $\RDDst$ constraint inferred from the DM method rather than from the HFLAV average (bottom-right panel). We refer to eq.~(\ref{eq:def-NP-shifts}) for WET WC notation, and to the text (\refsec{sec:SMEFT}, 2nd paragraph) for the meaning of $C_{ii23}$.}
  \label{fig:fit-RDDst}
  \end{center}
\end{figure}
The above scenario can thus be realized in different declinations, that we explore in \reffig{fig:fit-RDDst}. We switch on $C_{2223}$ and $C_{3323}$ independently. Through RG running, {\em each} of them would induce a contribution to $\delta C_9^{\uemu}$. However, $C_{2223}$ also generates a matching contribution to $\delta C_{LL}^{(\mu)}$, which is now constrained to a SM-like value. On the other hand, the $\delta C_9^{\uemu}$ shift induced by $C_{3323}$ is much less constrained and can be numerically dominant. Quite remarkably, fixing $C_{3323}$ to account for $\RDDst$ yields a $\delta C_9^{\uemu}$ shift of precisely the correct size to account for, {\em separately}, $b \to s \mumu$ BR measurements and $b \to s \mumu$ angular analyses. These facts are shown in the top-left panel of \reffig{fig:fit-RDDst}. In this panel, we assume $C_{1123} = 0$. In fact, a non-zero $C_{1123}$ would generate a matching contribution to $\delta C_{LL}^{(e)}$, that (in order to fulfil the $\RKKst$ constraint) one would want to be of similar size as $\delta C_{LL}^{(\mu)}$, that in turn is constrained to be nearly zero because of the $\Bsmm$ constraint. In short, the above scenario corresponds to $|C_{3323}| \gg |C_{2223}| >0$, and $C_{1123} = 0$, which can be justified on grounds of hierarchical NP.

Data, however, are also compatible with the alternative scenario $C_{2223} = C_{1123}$, with $|C_{3323}|$ again hierarchically larger in magnitude. This scenario can be justified on grounds of light-lepton universality, and is displayed in the top-right panel of \reffig{fig:fit-RDDst}, in the same WC plane as the top-left panel. We see that the main difference is the impact of the $\RKKst$ constraint, which becomes ineffectual in the top-right panel. This situation thus parallels that of the bottom-right vs. bottom-left panels of \reffig{fig:fit-9vs10}, and was already commented upon before the final items on page \pageref{page:two-messages}.

In the bottom-left panel of \reffig{fig:fit-RDDst} we show the implications of the same scenario ($|C_{3323}| \gg |C_{2223} = C_{1123}|$) in a different plane of WET WC combinations, that we consider more appropriate to capture the most promising effects in the light of present data. The $y$ axis corresponds to the left-handed WC shift relevant for $\RDDst$, namely $\delta C_{LL}^{(\tau)}$. In turn, the $x$ axis is the {\em total} (i.e. lepton-specific, matching-induced plus lepton-universal, RG-induced) $C_9$ shift, $\delta C_9^{\uemu} + \delta C_9^{(\mu)} = \delta C_9^{\uemu} + \delta C_9^{(e)}$, which is identical for both light leptons in view of the assumptions in the underlying SMEFT scenario.

The top-right vs. bottom-left panels are very correlated with one another---the latter is basically a rotation plus a reflection of the former. Jointly, however, they show that while the {\em muonic} $\delta C_{LL}$ direction has to a large extent been ``trivialized'' by $\Bsmm$, the {\em tauonic} $\delta C_{LL}$ direction is the now-promising one, in the light of a SMEFT interpretation of the $\RDDst$ discrepancy.

We emphasize that the SMEFT scenarios behind the top-left panel on one side and the top-right \& bottom-left panels on the other side are distinct. The fact that they yield mutually consistent results is non-trivial, and is due to the lack of constraining power in electronic-channel BR and angular data. Basically, the strongest constraint on $b \to s e^+ e^-$ at present is the one inferred indirectly from $\RKKst$. The day channel-specific data, i.e. $b \to s e^+ e^-$ BR and angular ones, will start to be constraining, they will either provide a striking confirmation of the above coherent picture, or expose an inconsistency. This inconsistency will signal that {\em (a)} the anomalous $\RDDst$ measurements, {\em (b)} the anomalous $b \to s \mumu$ BR ones, {\em (c)} the anomalous $b \to s \mumu$ angular ones, and {\em (d)} the electronic counterparts (whether also anomalous or not) of sets {\em (b) + (c)} are not consistent with a SMEFT picture. For reference, the counterparts of the top-left and top-right panels, but in the plane of the underlying SMEFT coefficients, are reported in the \hyperref[app:SMEFT]{Appendix}.

The above findings show once again that the wealth of semileptonic $b$-quark decay data offer a unique probe into possible heavy beyond-SM dynamics, and that this dynamics fulfils at the moment the basic constraints imposed by the SMEFT. We are lucky enough that this picture will be corroborated or falsified soon by data.
We meanwhile reiterate that a cautious approach to the above findings is in order. In this spirit, we also show in the bottom-right panel the counterpart of the bottom-left-panel plot, but for the fact that the $\RDDst$ best-fit region is inferred from the Dispersive-Matrix (DM) method \cite{Martinelli:2021frl, Martinelli:2021onb, Martinelli:2021myh}, rather than from the HFLAV average, whereby the latter represents the default choice in the rest of our numerical study. The DM method suggests a much milder $\RDDst$ anomaly than it emerges from the HFLAV average.  The bottom-right panel of \reffig{fig:fit-RDDst} thus allows to address quantitatively the question to what extent a non-zero tauonic $\delta C_{LL}$ gets closer to zero should the $\RDDst$ discrepancy fade to the below-$2\sigma$ figure suggested by the DM method.

For the sake of the discussion to follow, we focus on the scenario with a NP shift that is LFU in {\it both} the $C_9$ and $C_{10}$ directions, shown in the bottom-left panel of Fig.~\ref{fig:fit-9vs10}. Because LFU is imposed on both axes the constraining power of \RKKst vanishes, and the discrepant data shows a very clear departure from the SM expectation, with a substantial improvement over the SM hypothesis, and an agreement between the different sets of data at the 1$\sigma$ level. The best-fit intervals for this scenario are reported in \reftab{tab:NP-scenarios} (see first entry), and suggest
a 20\% NP effect in $\Cnine^{u(e,\mu)}$, whereas shifts in $C_{10}^{u(e,\mu)}$ stay consistent with zero. We take this case as one of our NP benchmarks, referred to as the ``real $\delta C_{9,10}$'' scenario in the following. The rest of the NP benchmarks in \reftab{tab:NP-scenarios} concern complex Wilson coefficients, to which we turn next.

\begin{table}[ht]
  \renewcommand{\arraystretch}{1.2}
  \centering
  \begin{tabular}{cccc}
    \toprule
    Scenario & Best-fit point & $1\sigma$ Interval & $\sqrt{\chi^{2, {\rm SM}} - \chi^2}$ \\
    \hline
$(\delta\Cnine^{u(e,\mu)}, \delta\Cten^{u(e,\mu)}) \in \mathbb{R}$ & $(-0.88,+0.30)$ & $([-1.08,  -0.56], [ 0.15 ,  0.46 ])$ & 5.5 \\
$\delta C_{LL}^{u(e,\mu)}/2 \in \mathbb{C} $ & $-0.70 - 1.36i$ & $[-1.00,  -0.54] +i [ -1.77,  -0.54]$ & 5.8 \\
$\delta C_{9}^{u(e,\mu)} \in \mathbb{C} $ & $-1.08 + 0.10i$ & $[ -1.31,  -0.85] +i [ -0.70, +0.85] $ & 6.4 \\
$\delta C_{10}^{u(e,\mu)} \in \mathbb{C}$ & $+0.68 + 1.40i$ & $[ +0.38,  +1.00 ] +i [ +0.69,  +1.92]$ & 3.2 \\
    \bottomrule
  \end{tabular}
  \caption{Reference scenarios for two-real or one-complex WC combinations.}
  \label{tab:NP-scenarios}
  \renewcommand{\arraystretch}{1.0} 
\end{table}

\Subsection{Complex Wilson-coefficient shifts}\label{sec:complex_WCs}

\begin{figure}[th!]
  \begin{center}
  \includegraphics[width=0.45\linewidth]{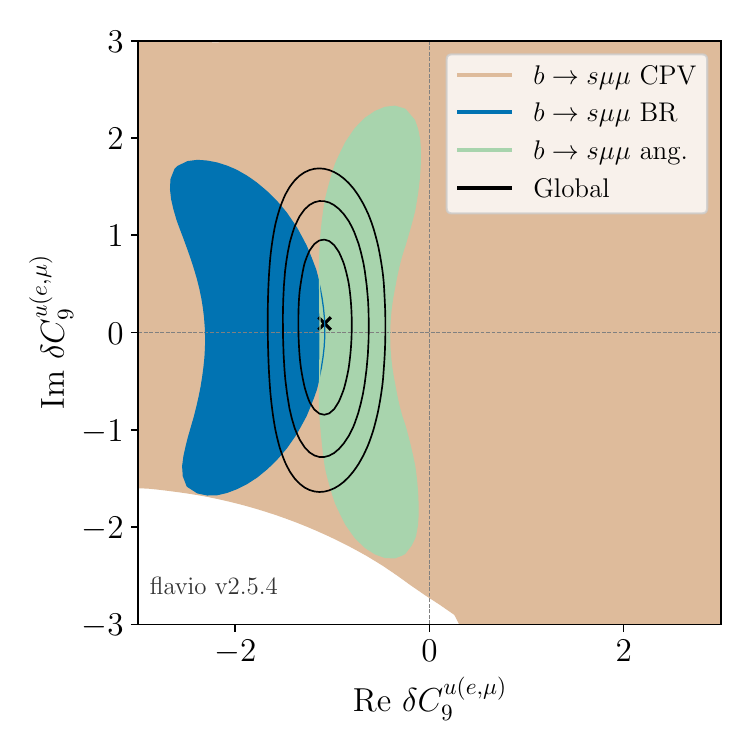}
  \includegraphics[width=0.45\linewidth]{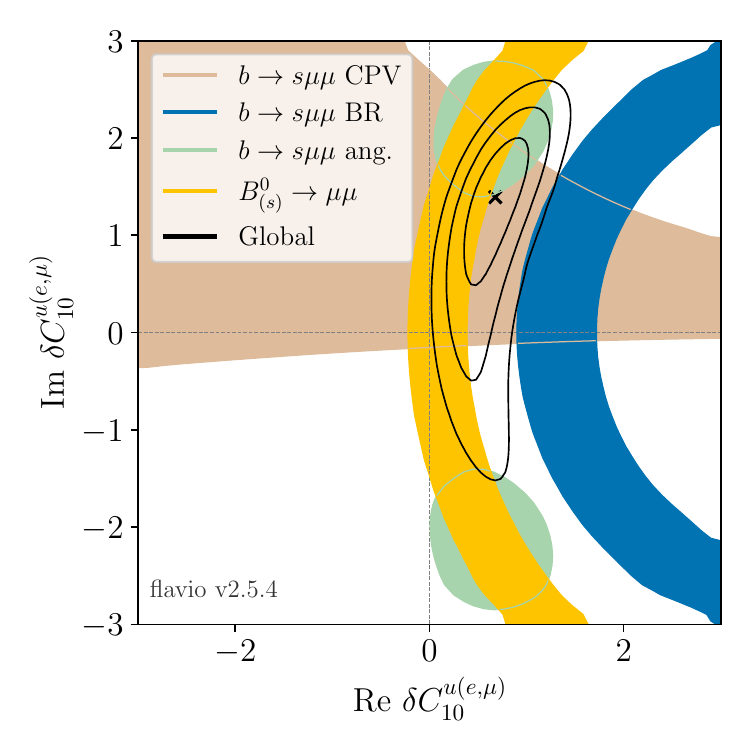}
  \includegraphics[width=0.45\linewidth]{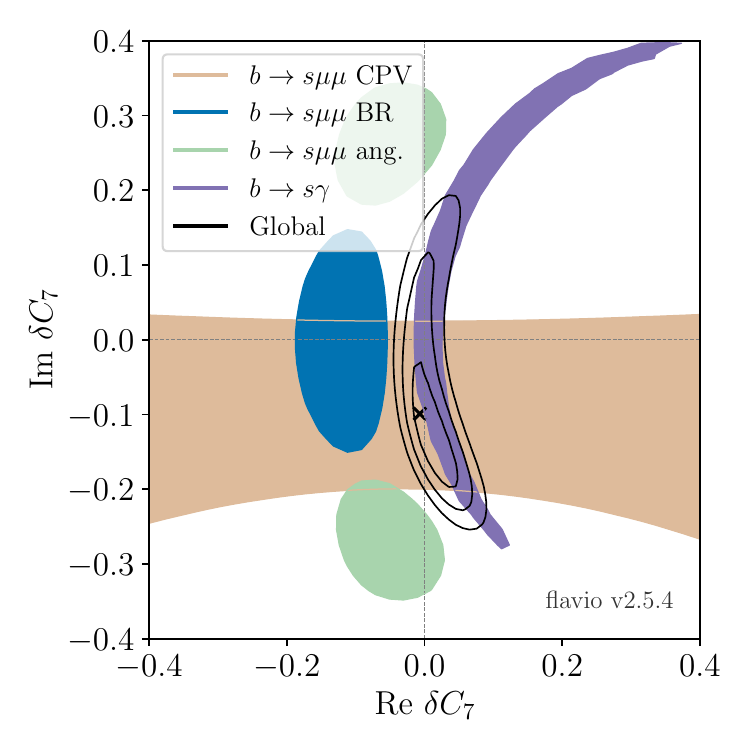}
  \includegraphics[width=0.45\linewidth]{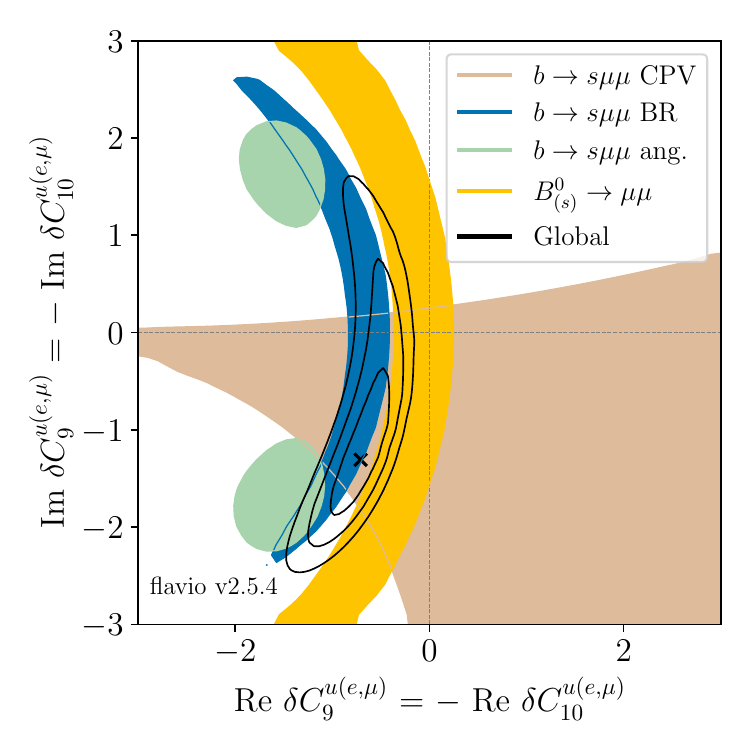}
  \caption{Global fits to the \btosmm data in the (top left) $\delta C_9^{\uemu}$ scenario, (top right) $\delta C_{10}^{\uemu}$ scenario, (bottom left) $\delta C_7$ scenario and (bottom right) $\delta C_{LL}^{\uemu}$ scenario.}
  \label{fig:complex-WC-fit}
    \end{center}
\end{figure}

It has been known for some time that the constraint on the magnitude of Wilson-coefficient shifts is loosened if their phase is not aligned with the phase of the SM contribution \cite{Altmannshofer:2012az}.\footnote{\label{foot:CP}This fact is clearly visible in global fits (see e.g. \cite{Carvunis:2021jga}), and is largely a consequence of the small number and limited accuracy of CP-odd observables available. Identifying scenarios with non-SM-like CP violation is interesting in its own right, given that new CP violation is, generally, a key requirement in the understanding of the composition of our universe, and given that flavour observables are prime probes of CP violation.} We quantify this possibility, i.e. consider scenarios where Wilson-coefficient shifts are complex, with generic phases. We do so by alternatively including or excluding CPV observables in the \btosmm sector, in order to study their possible role in favoring a particular sign for the imaginary part of the Wilson coefficients. Concretely, we include the e.m.-dipole Wilson coefficient
\be
\label{eq:C7_def}
C_7^{bs} - C_7^{bs,\rm SM} \equiv \delta C_7^{bs} \equiv \delta C_7~,~~~~~~~
C_7^{bs\{,\rm SM\}} \equiv C_7^{\{\rm SM\}}~
\ee
and explore four different scenarios, with a NP shift in one single Wilson coefficient or a combination, namely $\delta C_7$, or $\delta C_9^{u(e,\mu)}$, or $\delta C_{10}^{u(e,\mu)}$, or the previously defined $\delta C_{LL}^{u(e,\mu)}/2 = \delta C_9^{u(e,\mu)} = - \delta C_{10}^{u(e,\mu)}$.
In compliance with the \RKKst measurement, every scenario enforces LFU on both the real and imaginary parts of the Wilson coefficients. The results are presented in \reffig{fig:complex-WC-fit}, and the $1\sigma$ intervals for the scenarios that are most interesting for our purposes are also collected in \reftab{tab:NP-scenarios}.

We find that the current data are compatible with a sizeable ($\simeq$ few $10\% \times$ the size of the SM contribution, see footnote \ref{foot:WCs}) imaginary component in any of the $\delta C_9^{u(e,\mu)}$, $\delta C_{10}^{u(e,\mu)}$ and $\delta C_{LL}^{u(e,\mu)}$ scenarios in \reftab{tab:NP-scenarios}.
Also, none of the scenarios hints at a non-zero imaginary part, in all cases compatible with zero below the $2\sigma$ level. These conclusions, to be qualitatively expected, follow from the fact that CP-odd observables are still underconstraining, as already noted. The figure also shows that the only scenario where all considered subsets of data are in mutual agreement (i.e. all the colored subregions overlap somewhere in the plane) is the first scenario, where one fits to $\re\,\delta C_{9}^{u(e,\mu)}$ vs. $\im\,\delta C_{9}^{u(e,\mu)}$. This suggests, again, that $\delta C_{9}^{u(e,\mu)}$ is a necessary ingredient, and $\delta C_{10}^{(\mu)} = 0$ a preferred requirement, to achieve mutual agreement among all data.

The first three of the scenarios collected in \reftab{tab:NP-scenarios} perform comparably well, in particular $\delta C_9^{\uemu}$ or $\delta C_{LL}^{\uemu}$ represent the best-performing complex scenarios. Interestingly, the complex-$\delta C_{LL}^{\uemu}$ case shows that a sizeable {\it real} NP contribution---of $\mc O(20\%)$ the SM value---is still allowed by the current \Bsmm measurement. This is because $\mc B(\Bsmm) \propto |C_{10}^{(\mu)}|^2$, leading to an approximately circular shape in the $\re~ \delta C_{LL}^{u(e,\mu)}$ vs. $\im~ \delta C_{LL}^{u(e,\mu)}$ plane. Also interestingly, \Bsmm is likewise compatible with a sizeable imaginary part---that is entirely plausible, see footnote \ref{foot:CP}, and that existing data barely probe. As regards $\delta C_7$, the strongest constraint comes from the wealth of \btosg data available, plus CPV observables in the $b \to s \mumu$ sector. Taking also into account that \Bsmmy at high $q^2$ is mostly sensitive to $C_{9,10}$ (see discussion in \refsec{sec:Bsmmy_th}), we do not consider a $\delta C_7$ scenario in the rest of this work.

\medskip

As a final remark for Sec. \ref{sec:global-fits} we emphasize again, as we did in the Introduction, that a complete calculation of non-local contributions entering the currently discrepant $b \to s \mm$ data may, {\em or may not}, show that the shifts collected in \reftab{tab:NP-scenarios}, in particular those involving the muonic $C_9$, are actually due to SM long-distance dynamics. In these circumstances, it is meaningful to pursue observables that bear sensitivity to the very same Wilson-coefficient shifts one wants to probe, but are {\em not} affected by the same long-distance physics. The \Bsmmy branching ratio at high $q^2$ is one such observable. In the following we will discuss the significance one may expect for a shift as large as the real-$\delta C_{9,10}$ scenario (first entry of \reftab{tab:NP-scenarios}) or the complex-$\delta C_{LL}$ one (second entry of the same table) through a high-$q^2$ analysis of $\mc B(\Bsmmy)$, as a function of the data accumulated.

\Section{$\Bsmmy$ Uncertainties and NP Reach} \label{sec:Bsmmy-exp-SM-NP}

\Subsection{Experimental uncertainties}\label{sec:Bsmmy_exp}

In great synthesis, measuring \Bsmmy at LHC using the partially reconstructed method \cite{Dettori:2016zff} relies on a fit to the $q^2$-differential distribution in an appropriate high-$q^2$ region, using as external constraints the known BRs of the purely leptonic modes $B_{d,s} \to \mm$ as well as the shapes and normalizations for certain well-defined backgrounds. It is difficult to assess the sensitivity of such procedure to \Bsmmy in the absence of a search optimized for \Bsmmy, rather than for the leptonic modes, as is the case for existing analyses.

With this important caveat in mind, the aim of the present section is to assess to the best of our knowledge the LHC prospects of measuring the \Bsmmy observable defined above. To this end, we need to make certain assumptions spelled out next. The first is that the efficiency of \Bsmmy in the considered integrated high-$q^2$ region is equal to \Bsmm's. This assumption relies on two opposing effects. On the one side, because \Bsmmy is not reconstructed as a peak but as a shoulder within a broad\footnote{As compared with the \Bsmm $q^2$ ``window'' namely the region where most of the \Bsmm signal candidates are.} dimuon invariant mass window, the efficiency will have a certain $q^2$ dependence. Typically the efficiency will decrease moving from the $B^0_s$ mass to lower $q^2$ values.
In fact, lowering $q^2$ means a lower transverse momentum in the laboratory frame thus lowering the reconstruction and triggering efficiencies. It also means more abundant backgrounds, from more channels, to be rejected with tighter requirements.
On the other side, this effect could be mitigated by developing an analysis optimised for the \Bsmmy decay, which Refs.~\cite{LHCb:2021vsc,LHCb:2021awg} are not.
More precisely, a dedicated \Bsmmy analysis could mitigate the $q^2$ dependence of the efficiency by deploying techniques, specific to partially-reconstructed decays, that were not necessary for the aims of Refs.~\cite{LHCb:2021vsc,LHCb:2021awg}. 
Explicitly, two dominant sources of background, in addition to the combinatorial one, are $B \to h \mu \nu$ decays, with $h$ being a hadron mis-identified as a muon, and $B \to \pi^0 \mm$, with a $\pi^0$ not reconstructed. Both decay sources need to be constrained in terms of the dimuon mass distribution as well as of the yield, in order not to limit the sensitivity to the signal. The improvements required are in the calibration of the trigger, and of the reconstruction and selection efficiencies as well as in particle (mis)-identification for muons (hadrons). These improvements are impossible to quantify reliably without a dedicated experimental study \cite{Camille's-Thesis}.

Clearly, since the background increases for smaller dimuon masses, the lower bound, $q^2_{\rm min}$, of $q^2$ integration, has to be chosen as a compromise between larger statistics and larger background.
In addition below a certain  $q^2_{\rm min}$ value around $(4~\GeV)^2$ $c\bar c$ resonances start to play an important role, lessening experimental and theoretical sensitivities alike. 
Within a given $q^2_{\rm min}$ choice, the actual sensitivity will rely on controlling the background within an associated error which is better than the expected signal yield. 
For example, the expected \Bsmmy events in the analysis of Ref.~\cite{LHCb:2021awg}, were $N_{\rm exp}(\Bsmmy) = 1.7$ with the full Run-2 dataset, for $q^2_{\rm min} = 4.9$ GeV$^2$ (table 7 of Ref.~\cite{LHCb:2021awg}) and a multivariate analysis BDT score $> 0.25$, meaning that the efficiency corresponding to the BDT requirement alone is 75\%.\footnote{Note that, to obtain this result, the BDT is trained with \Bsmm as signal, i.e. is {\em not} optimized for \Bsmmy instead: from e.g. a BDT score of 1, meaning a maximal \Bsmm purity, one cannot unambiguously infer the \Bsmmy purity.} This $N_{\rm exp}(\Bsmmy)$ figure may be compared with the {\em accuracy} to which the leading background is controlled. From table 9 of Ref.~\cite{LHCb:2021awg}, this accuracy is 5 events with the same BDT requirement. This comparison suggests that background calibration has clearly to improve, but it leaves hope, because the two numbers (1.7 events of signal on a background uncertainty of 5) are not far from each other even within an analysis far from optimized for our signal of interest. It is important to note that the latter argument considers the expected yields and their uncertainties integrated over the {\it entire mass window}, while a fit to the dimuon mass distribution gives a superior sensitivity provided the shapes of the background contributions are known. 

In the light of the above considerations, we will henceforth assume that all the backgrounds are under control, i.e. that their uncertainties will eventually fall safely below the signal yield. Under this ``no-background'' hypothesis the \Bsmmy-signal uncertainty is dominated by the sheer amount of data collected. We accordingly assess the \Bsmmy sensitivity to the NP scenarios in \reftab{tab:NP-scenarios} as a function of the data size. Finally, while what follows assumes the LHCb experiment as reference, the same techniques can be used also in other setups where we can expect copious amounts of $B_s$ mesons to be produced, e.g. at ATLAS and CMS. The level of the combinatorial background will however depend on the interactions pile-up and the control of the misidentified background will rely on details of the particle identification of each experiment. 

\Subsection{Theory uncertainties} \label{sec:Bsmmy_th}

A reappraisal of the theoretical uncertainty in $\mc B(\Bsmmy)$ for high $q^2$ above narrow charmonium was presented in Ref.~\cite{Guadagnoli:2023zym}. It was shown that the main theory uncertainty arises from the vector and axial form factors $V_{\perp,\parallel}(q^2)$ of the \Bstog transition; that tensor form factors $T_{\perp,\parallel}(q^2)$ play, in comparison with vector and axial ones, a subdominant role irrespective of the parameterization used; that broad charmonium plays an entirely negligible role for $\sqrt{q^2} \gtrsim 4.2$ GeV. With the existing estimations of the vector and axial form factors---none of which is based exclusively on a first-principle calculation---the uncertainty in $\mc B(\Bsmmy)$ integrated in $\sqrt{q^2} \in [4.2, 5.0]$~{\GeV} is currently on the order of 50\%. To resolve $\delta C_9^{(\mu)} / C_9^{(\mu), \rm SM} \simeq 15\%$, one has to control the vector and axial form factors multiplying this Wilson coefficient to an accuracy better than $\delta C_9^{(\mu)} / C_9^{(\mu), \rm SM}$, because the BR has the {\em same}, dominantly quadratic dependence on {\em both} $C_9^{(\mu)}$ and the vector and axial form factors. Controlling form factors with such accuracy is {\em already} within reach of LQCD calculations. Specifically, Ref.~\cite{Desiderio:2020oej} calculates $D_s \to \gamma$ vector and axial form factors with a quoted error around 10\% or even below. An even higher precision has been achieved in the very recent Ref.\,\cite{Frezzotti:2023ygt}, where some lattice data reach a relative error of few percent. Besides, these form factors are computed {\em directly} in the very same high-$q^2$ region of interest for the indirect $\mc B(\Bsmmy)$ measurement, i.e. no kinematic extrapolation is required.\footnote{Given the decay $P(p) \to \mu\mu(q) + \gamma(k)$, with $P$ the decaying meson, and introducing the variable $x_\gamma \equiv 2 pk / m_P^2 = 1 - q^2 / m_P^2$, Ref.~\cite{Desiderio:2020oej} calculates $D_s \to \gamma$ FFs for $x_\gamma \in [0.05, 0.4]$. Our fiducial $q^2$ range for \Bsmmy, $\sqrt{q^2} \in [4.2, 5.0]\,\GeV$, corresponds to $x_\gamma \in [0.39, 0.13]$.}

In short, a theory error as small as needed is well within reach to the extent that the calculation in Refs.~\cite{Desiderio:2020oej, Frezzotti:2023ygt} is extended to the $B_s$ case. Besides, it is realistic to assume that ripe LQCD calculations of $B_s \to \gamma$ vector and axial form factors in the very kinematic region of interest to us will come with typical errors not larger than 5\%, which is negligible in comparison with the nominal NP shift mentioned above and also with the experimental error to be expected in the foreseeable future.

\begin{figure}[th!]
  \begin{center}
    \includegraphics[width=0.495\linewidth]{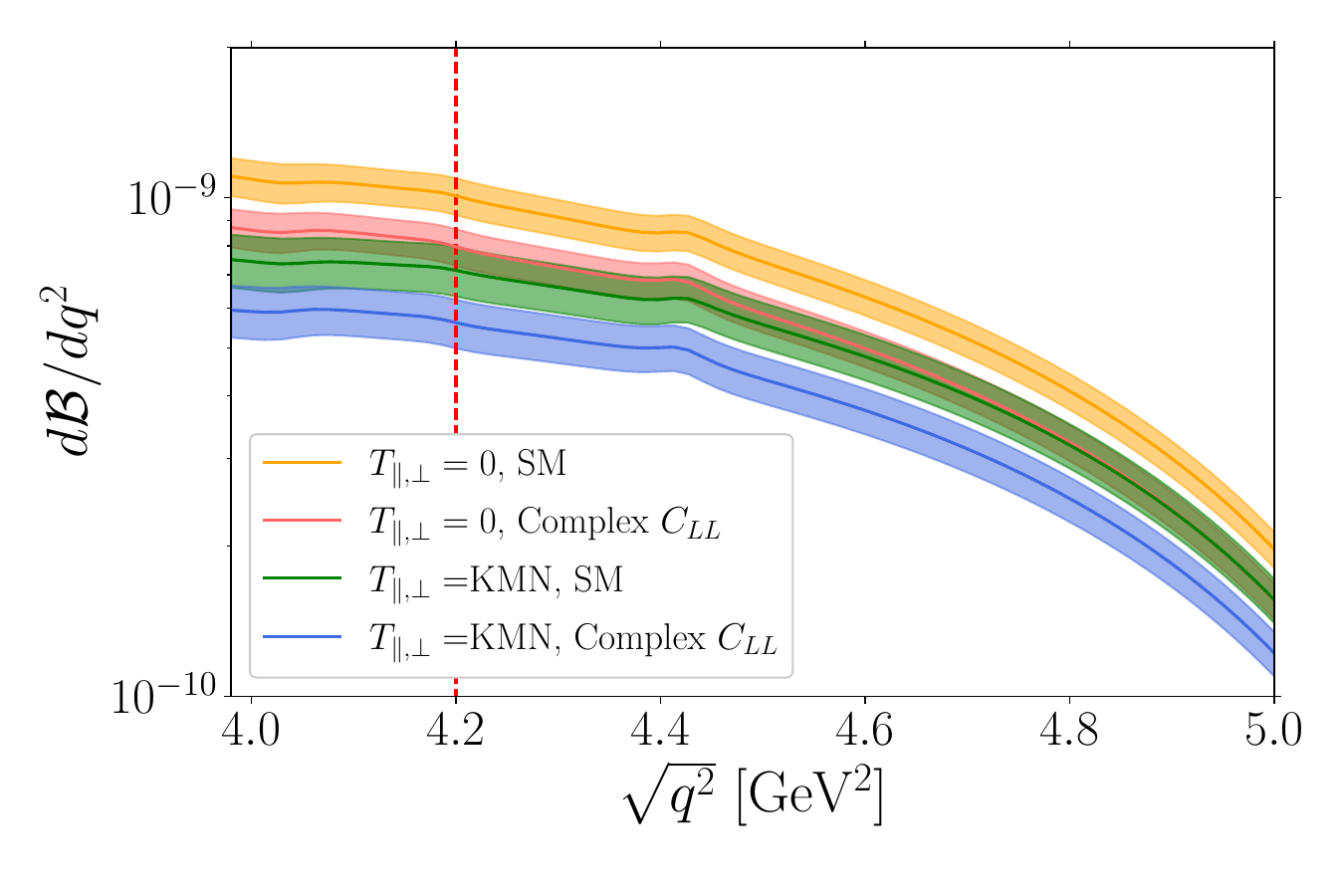}
    \includegraphics[width=0.495\linewidth]{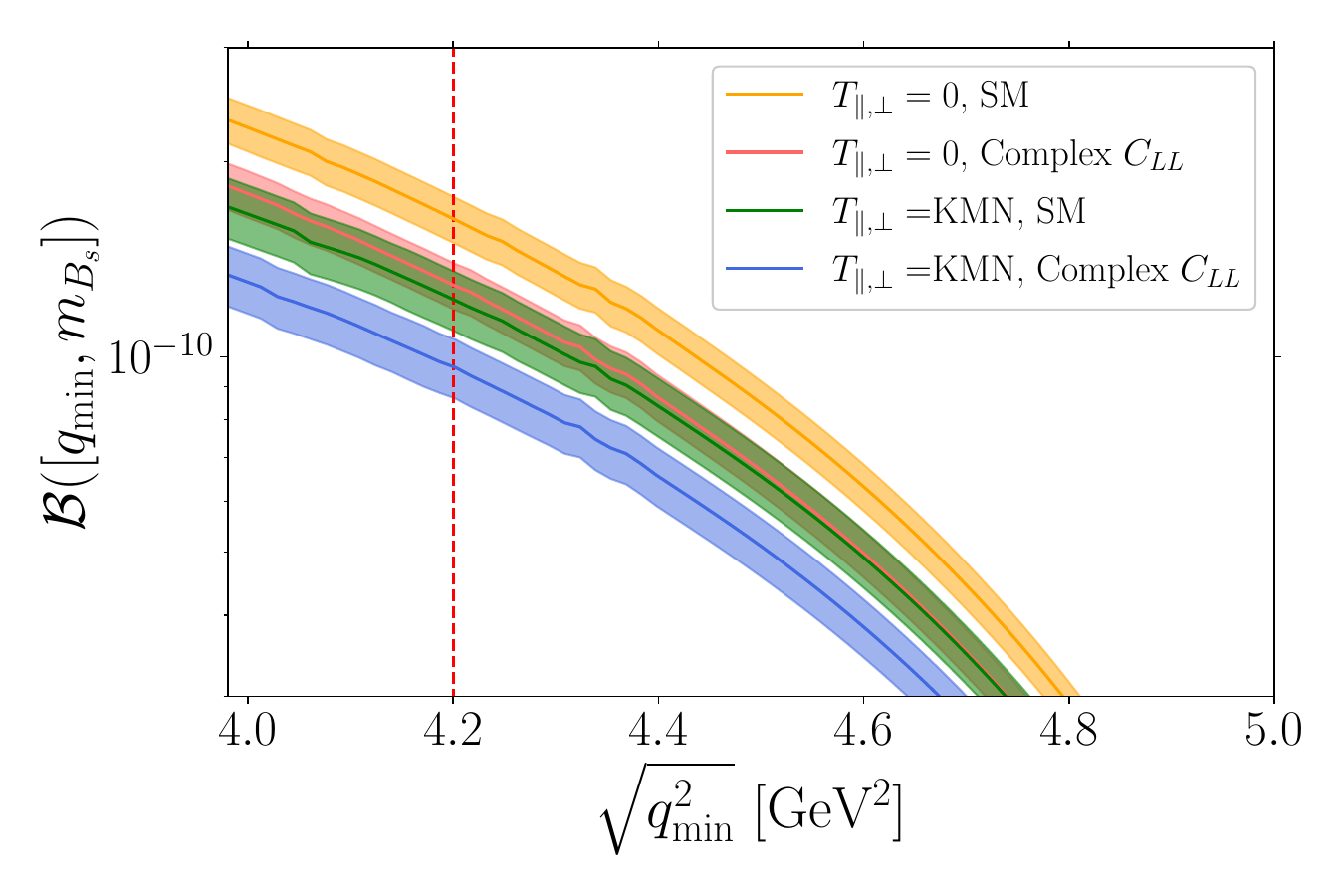}
  \caption{(left) Differential branching fraction in $q^2$ and (right) integrated branching ratio of \Bsmmy in the high-$q^2$ region as a function of the lower bound of integration $\sqrt{q_\textrm{min}^2}$, for various parameterizations of the tensor form factors $T_{\parallel, \perp}$ and two relevant theory scenarios. A 5\% uncertainty is assumed for the vector and axial form factors, with central values from Ref.~\cite{Guadagnoli:2023zym}. The central values for the $T_{\parallel, \perp}$ form factors labelled as KMN are taken from Ref.~\cite{Kozachuk:2017mdk}. They are shown with a 20\% uncertainty, following the discussion end of Sec.~\ref{sec:Bsmmy_th}. The $C_7$ Wilson coefficient is floated within the $1\sigma$ region allowed by the complex fit of \reffig{fig:complex-WC-fit}.}
  \label{fig:BR-FTVA}
    \end{center}
\end{figure}

For the purpose of the present study we thus mostly need realistic central values for the vector and axial $B_s \to \gamma$ form factors, and we adopt the recent parametrization proposed in Ref.~\cite{Guadagnoli:2023zym}. In that work, the contributions from tensor and axial-tensor form factors were found to be negligible within the still large uncertainty on the vector and axial-vector form factors obtained in the same work. However, in the present case the vector and axial form-factor uncertainty is assumed to be below 5\%, as already discussed, and the choice of tensor form factors does have an impact, as can be seen in \reffig{fig:BR-FTVA}. For example, one may compare the parameterization in Ref.~\cite{Kozachuk:2017mdk} with setting $\FTVA=0$, which means assuming a 100\% uncertainty. The difference between the two choices induces a difference in the BR prediction of the {\em same order} as the NP shift one wants to access.

\begin{figure}[th!]
  \begin{center}
    \includegraphics[width=0.495\linewidth]{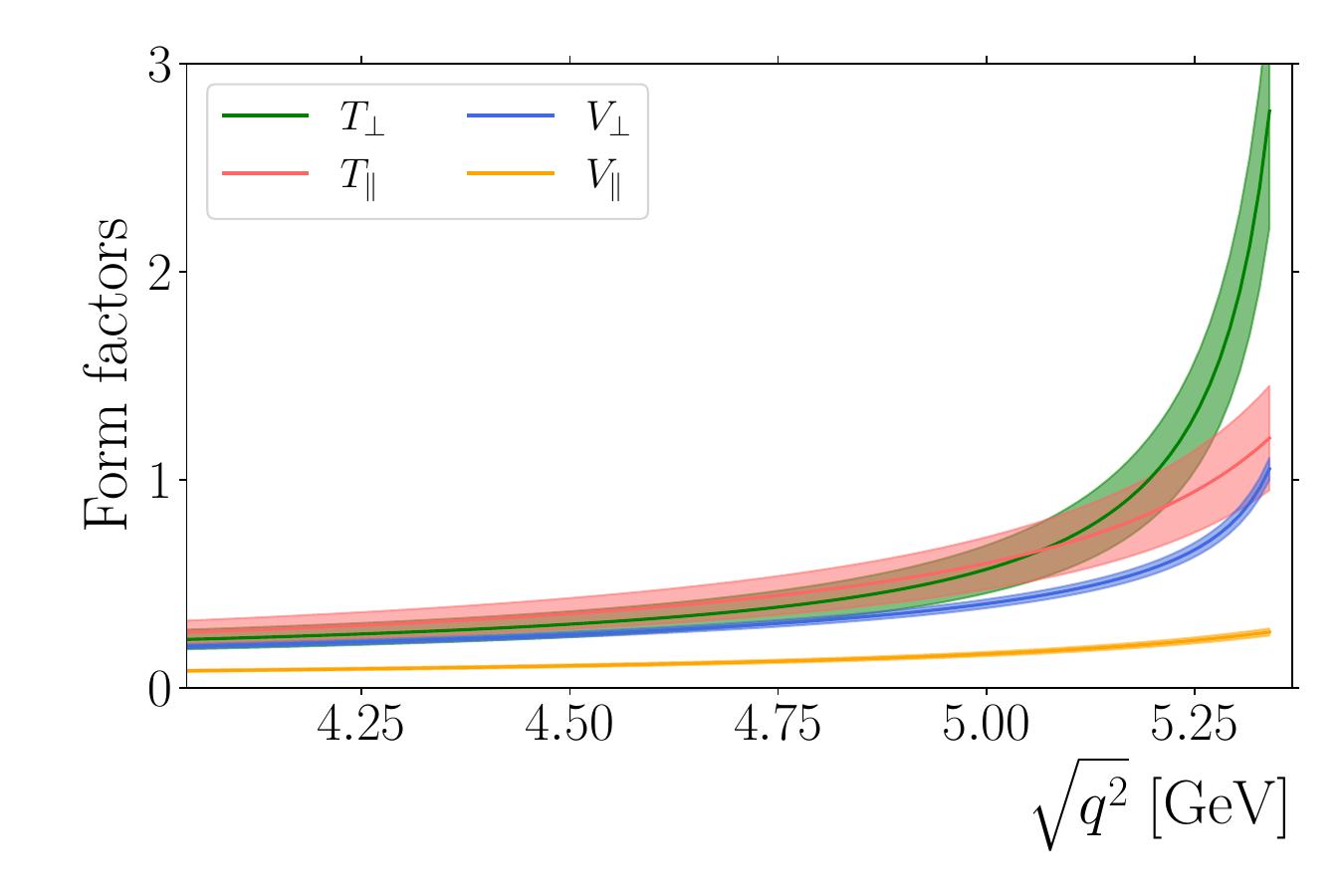}
    \caption{Set of \Bstog transition form factors. For the assumed $V_{\parallel,\perp}$ and $T_{\parallel,\perp}$ uncertainties, see considerations in the caption of \reffig{fig:BR-FTVA} and end of Sec.~\ref{sec:Bsmmy_th}.
    }
  \label{fig:FTVA}
    \end{center}
\end{figure}

In short, a first-principle evaluation of the tensor form factors will also be required. But importantly, a limited accuracy, of the order 20\%, will be {\em sufficient} in this case, given the subdominant nature of these contributions. In fact, a 20\% error on tensor form factors, which appear in contributions whose overall nominal size does not exceed about 25\% of those induced by vector or axial form factors, is sufficient to control tensor contributions with an overall theory uncertainty of about 5\%, which is already below the size of the expected NP shift and the assumed uncertainty on the vector and axial form factors. The final uncertainties we assume for the form factors are shown in \reffig{fig:FTVA}.

As concerns the short-distance side of the tensor contributions, we vary $\delta C_7$ in the $1\sigma$ interval identified from the complex fit of \reffig{fig:complex-WC-fit}. We do not pursue a sensitivity study to possible NP contributions to $C_7$, in particular because \Bsmmy at high $q^2$ is vastly more sensitive to $\delta C_{9,10}^{(\mu)}$. Sensitivity to $C_7$, whose SM value is much smaller than the $C_{9,10}^{(\mu)}$ counterparts, requires to be close to the {\em lower} $q^2$ endpoint. Also, sensitivity to $C_7$ occurs through terms $\propto {\rm Re}(C_7 \, C_{9,10}^{(\mu)*})$ and is thus only linear. Indeed, the complex NP shift to $C_7$ shown in \reffig{fig:complex-WC-fit} is too small to be resolved.

\Subsection{Outlook on $\Bsmmy$ as a probe to NP}\label{sec:Bsmmy_NP}

In this section, we put to work the considerations made in Secs.~\ref{sec:Bsmmy_exp}-\ref{sec:Bsmmy_th} and use them to infer the NP sensitivity of a measurement of $\mc B(\Bsmmy)$ at high $q^2$ with the partially reconstructed method \cite{Dettori:2016zff}. For clarity, the main conclusions of Secs.~\ref{sec:Bsmmy_exp}-\ref{sec:Bsmmy_th} are that: {\em (i)}~we may assume a {\em theory} uncertainty on $\mc B(\Bsmmy)$ of about 10-12\%, dominated by FFs. This uncertainty can be understood intuitively, because FFs contribute quadratically to the BR, and their uncertainty is dominated by $V_{\perp,\parallel}$, that in our $q^2$ range of interest is determined with an accuracy of 5\% or less. As discussed in the previous section, other sources of theory uncertainties, including the $T_{\perp,\parallel}$ error, the dependence on $\delta C_7$, and broad charmonium, are subdominant;\footnote{These considerations are illustrated by the predictions in eq.~(\ref{eq:BR_3scenarios}) to follow.} {\em (ii)}~the theory error is actually negligible with respect to the experimental error. The latter is difficult to infer with any confidence. Hereafter, we consider the case where it is dominated by sheer statistics, which may be a safe and not unrealistic assumption (see Sec. \ref{sec:Bsmmy_exp} for details).

As regards the assumed NP shifts we focus on two scenarios, referred to as ``real $\delta C_{9,10}$'' and ``complex $\delta C_{LL}$'', and corresponding to the first two entries of \reftab{tab:NP-scenarios}. In either of the two cases, the shift is assumed to be  LFU,\footnote{In principle, the minimal assumption is light-lepton universality, as highlighted by the $^{\uemu}$ index in \reftab{tab:NP-scenarios}. However, the case of full lepton universality, denoted as LFU without further qualifications, is not currently distinguishable from the former.} hence we drop the flavour index hereafter. Note that we do not consider other logically possible scenarios, to which the \Bsmmy sensitivity is either too small---e.g. scenarios involving $C_7$---or is similar to the real-$\delta C_{9,10}$ and complex-$\delta C_{LL}$ scenarios we focus on.

Using the best-fit values in \reftab{tab:NP-scenarios}, and the theoretical uncertainties discussed in Sec.~\ref{sec:Bsmmy_th} and summarized above, we extract the integrated BR for \Bsmmy in eq.~(\ref{eq:BR_3scenarios}). A qualification is in order about the chosen $q^2$ range. The photon in \Bsmmy is to be understood as ``Initial State Radiation''. Then, as discussed in Refs.~\cite{Dettori:2016zff,Guadagnoli:2023zym}, $\mc B(\Bsmmy)$ is well-defined for $\sqrt{q^2} < 5$ GeV, because above this threshold ``Final State Radiation'' (FSR), or bremsstrahlung from the muons, is not negligible or dominant. However, experimentally \Bsmmy and \Bsmm are two components of the same fit; di-muon candidates in this fit are bremsstrahlung recovered, i.e. FSR is in practice subtracted by MonteCarlo \cite{Davidson:2010ew}; finally, choosing $[4.2, 5.0]~\GeV$ or $[4.2~\GeV, m_{B_s^0}]$ leads to a difference in the prediction by barely 2\%. Hence, following Ref.~\cite{Guadagnoli:2023zym}, the predictions in eq.~(\ref{eq:BR_3scenarios}) refer to $\sqrt{q^2} \in [4.2~\GeV, m_{B_s^0}]$.
\begin{eqnarray}
\label{eq:BR_3scenarios}
\textrm{SM}	        &:& ~ (1.22  \pm 0.12_{(\FV,\FA)} \pm 0.06_{(\FTV,\FTA)} \pm 0.04_\textrm{other}) \times 10^{-10} ~, \nn \\
\mbox{real $\delta C_{9,10}$}   &:& ~ (0.90  \pm 0.09_{(\FV,\FA)} \pm 0.05_{(\FTV,\FTA)} \pm 0.03_\textrm{other}) \times 10^{-10} ~, \\
\mbox{complex $\delta C_{LL}$}  &:& ~ (0.99  \pm  0.10_{(\FV,\FA)} \pm 0.04_{(\FTV,\FTA)} \pm 0.05_\textrm{other}) \times 10^{-10} ~.\nn
\end{eqnarray}
We emphasize that the quoted uncertainty is theoretical only, and takes into account all other known sources, including $V_{cb}$ and the broad-charmonium resonances\footnote{We refer the reader to Ref.~\cite{Guadagnoli:2023zym} for details about this aspect.} and also marginalizing over $\delta C_7$ in the NP scenarios (lines 2 and 3 of eq.~(\ref{eq:BR_3scenarios})). The central values in the second and third lines represent 20\% shifts compared to the SM expectation (first line). The latter matches the estimate from Ref.~\cite{Guadagnoli:2023zym}, where central values for tensor FFs are taken from Ref.\,\cite{Kozachuk:2017mdk}.

\begin{figure}[th!]
  \begin{center}
    \includegraphics[width=0.45\linewidth]{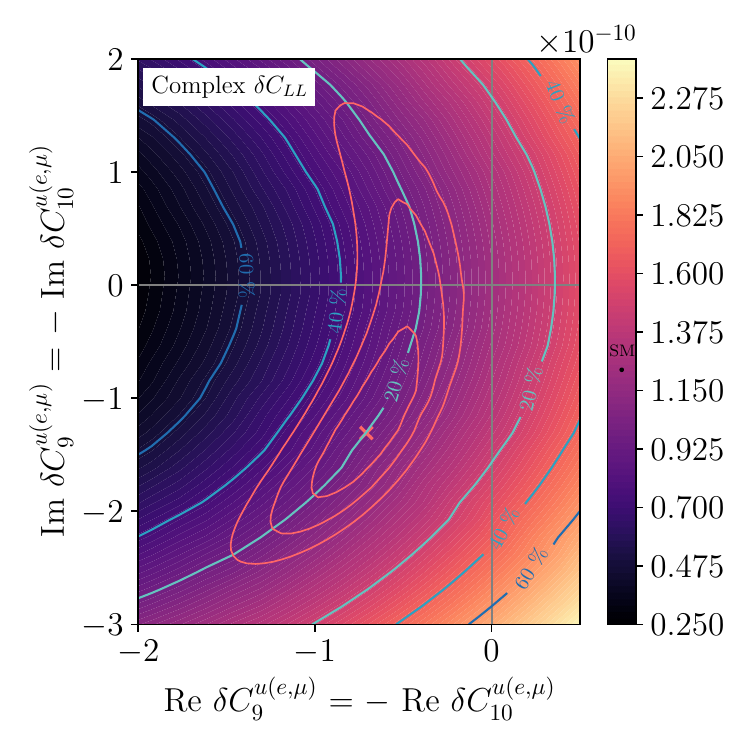}
    \includegraphics[width=0.45\linewidth]{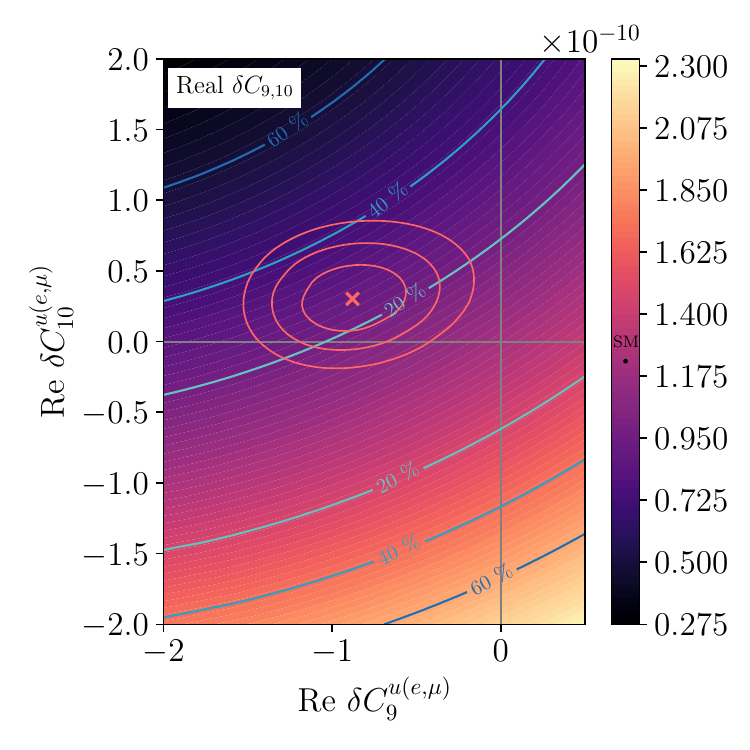}
  \caption{Integrated branching ratio of \Bsmmy in the high-$q^2$ region, (left) in the plane $\re \, C_{9}^{u(e,\mu)}  = - \re \, C_{10}^{u(e,\mu)}$ vs. $\im \, C_{9}^{u(e,\mu)} = - \im \, C_{10}^{u(e,\mu)}$, i.e. in the complex-$\delta C_{LL}$ scenario, and (right) in the plane $\re \, C_{9}^{u(e,\mu)}$ vs. $\re \, C_{10}^{u(e,\mu)}$, i.e. in the real-$\delta C_{9,10}$ scenario.}
  \label{fig:BR-vs-WCs}
    \end{center}
\end{figure}
\begin{figure}[th!]
  \begin{center}
    \includegraphics[width=0.45\linewidth]{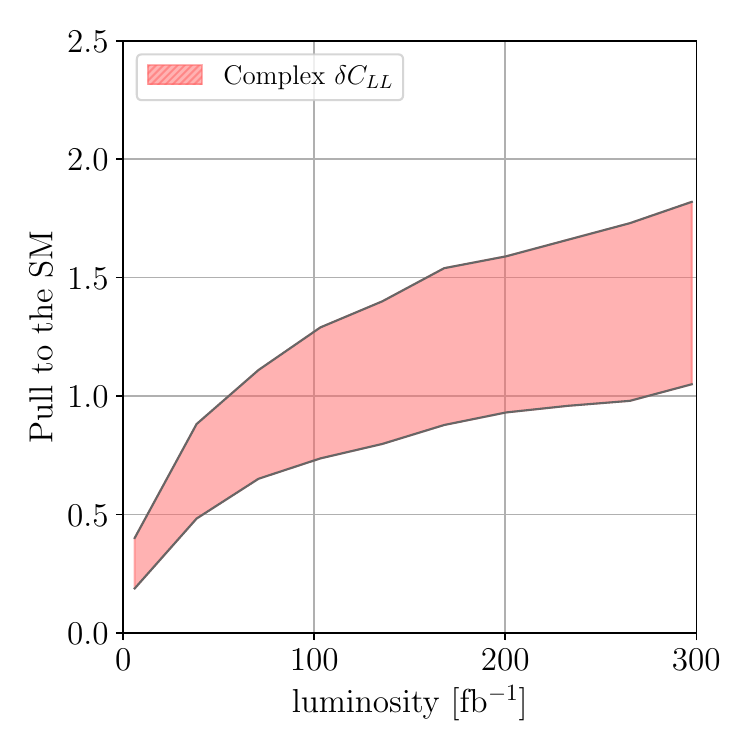}
    \includegraphics[width=0.45\linewidth]{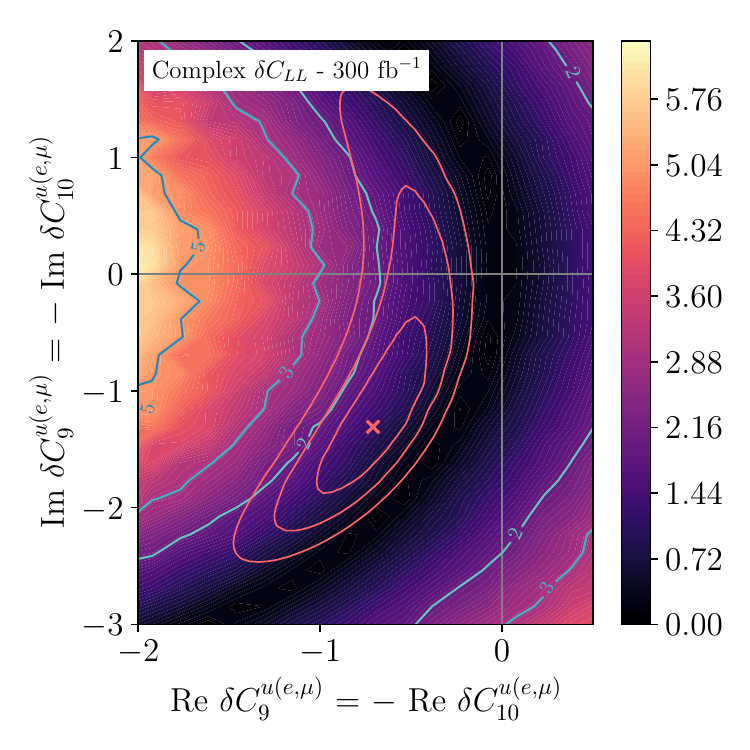}
    \includegraphics[width=0.45\linewidth]{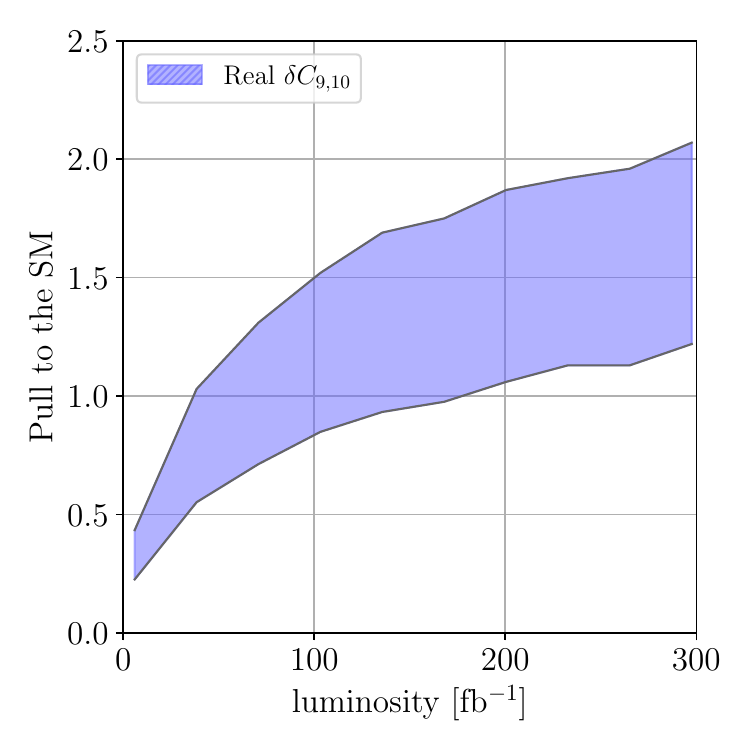}
    \includegraphics[width=0.45\linewidth]{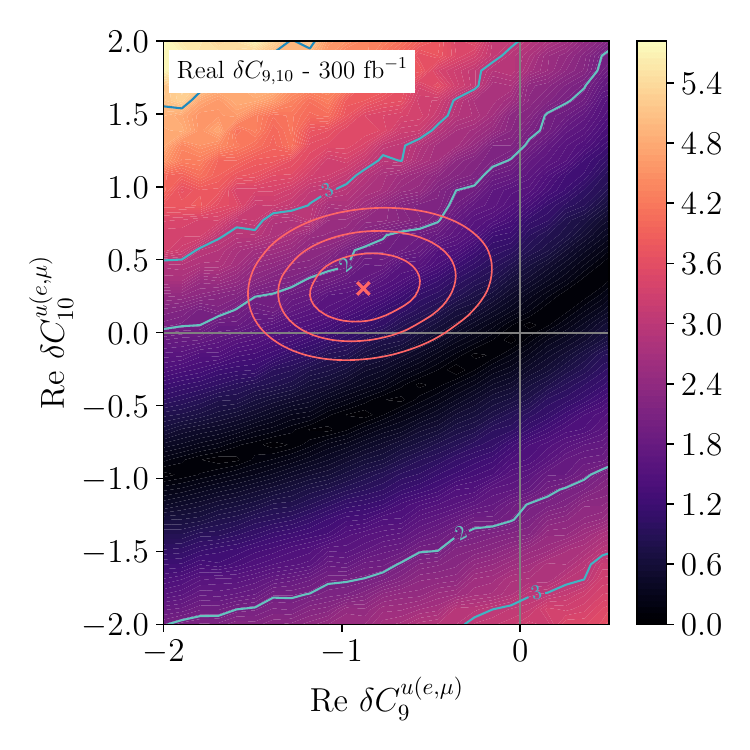}
  \caption{Pull to the SM (top) in the complex $\delta C_{LL}$ scenario and (bottom) in the real $\delta C_{9,10}$ scenario. The colored area on the leftmost panels represents the $1\sigma$ region spanned by the respective NP scenarios. The rightmost panels show the pull for $300\invfb$ of collected data.}
  \label{fig:pull-vs-lumi}
    \end{center}
\end{figure}

The sensitivity of \Bsmmy in the high-$q^2$ region to the two NP scenarios is displayed in \reffig{fig:BR-vs-WCs}. The leftmost panel shows that $\mc B(\Bsmmy)$ provides stronger constraints on the real part of $\delta C_{LL}$, as expected from a CP-even observable. Using this NP benchmark (i.e. a $\delta C_{LL}$  leading to the last of eqs.~(\ref{eq:BR_3scenarios})) as the central-value prediction for the integrated branching ratio, one can compute the relative error on the \Bsmmy measurement. The pull of this observable to the SM is shown in \reffig{fig:pull-vs-lumi}.
We note that, for each luminosity value in the figure, the range in the pull corresponds to the $1\sigma$ range in the Wilson-coefficient shift.
In turn, the sensitivity of \Bsmmy to the real-$\delta C_{9,10}$ scenario is shown in the rightmost panel of \reffig{fig:BR-vs-WCs}. We see that sensitivity to this scenario is enhanced compared to the complex-$\delta C_{LL}$ one. Even if the presence of NP {\em lowers} the integrated BR---and thus the event yield---w.r.t. the complex-$\delta C_{LL}$ case, we find that the pull to the SM in the real-$\delta C_{9,10}$ scenario is larger than the pull for the complex-$\delta C_{LL}$ one as a function of the acquired data and reaches the $2\sigma$ level at the border of the $1\sigma$ region for the Wilson-coefficient shift. We finally note that, although other scenarios in \reftab{tab:NP-scenarios} have a higher pull (see last table column), the BR for \Bsmmy has stronger constraining power in scenarios that involve the real parts of $C_9$ {\em and} $C_{10}$, if possible independently. Scenarios featuring imaginary shifts to Wilson coefficients perform better over the SM precisely because constraints from BRs are weaker. For such scenarios, one would have to resort to CP-sensitive observables, e.g. $A_{\Delta \Gamma}$ \cite{Carvunis:2021jga} in the case of our decay.

\Section{Conclusions}
\label{sec:concl}

Purpose of this paper is a quantitative study of the potential of $\Bsmmy$, measured at high $q^2$ as a partially reconstructed decay \cite{Dettori:2016zff}, as a probe of the origin of the discrepancies in semi-leptonic $b \to s$ and $b \to c$ decays.

As an initial step towards this end, we reassess these discrepancies, first in the weak effective theory, then in the SMEFT. Our global fit displays quantitative consistency across $b \to s$ and $b \to c$ semi-leptonic data---even after the \RKKst and \Bsmm updates. Specifically, \Bsmm makes the {\em muonic} $\delta C_9 = - \delta C_{10}$ direction less appealing than it used to be before the updates, because it prefers $C_{10}$ to be SM-like within errors. Instead, data provide circumstantial support to a {\em tauonic} $\delta C_9 = - \delta C_{10} \equiv C_{LL}/2$ shift generated at a high scale, and leaving as imprints effects in $R(D^{(*)})$ on the one side, {\em and} a lepton-universal $\delta C_9$ on the other side. SMEFT allows to make this connection between $b \to s$ and $b \to c$ effects quantitative. We find that the tauonic $C_{LL}$ required by $R(D^{(*)})$ implies a universal $\delta C_9$ of precisely the size required, {\em separately}, by the anomalous $b \to s \mumu$ BR data, and by the anomalous angular $b \to s \mm$ analyses. Needless to say, this conclusion is to be taken with great caution, because both imprints are far from established: first, angular, and especially BR data demand updates; second, the significance of the $R(D^{(*)})$ anomalies is in the eye of the global-fit beholder and ranges from $\sim 2\sigma$ \cite{Martinelli:2021frl, Martinelli:2021onb, Martinelli:2021myh} to higher significances \cite{HeavyFlavorAveragingGroup:2022wzx}, depending on the analysis strategy and/or on the theoretical inputs used.

The above-mentioned global fits allow us to identify reference scenarios, that we use as benchmarks for our stated aim of exploring the potential of $\Bsmmy$ at high $q^2$ as a probe of the flavour anomalies. For this purpose, we first discuss the outlook on the total error of this observable, from both theory and experimental standpoints. Importantly, the theory accuracy of $\mc B(\Bsmmy)$ is not limited by the long-distance effects that inherently hinder predictions for semi-leptonic $b \to s$ modes at low-$q^2$, for example effects due to $B \to D \bar {D}^*$ rescattering. In this respect, $\Bsmmy$ offers a neat strategy to probe the very same short-distance physics possibly responsible for the anomalies, but in a {\em different} kinematic region. The $\Bsmmy$ theory error at high $q^2$ is dominated by the form-factor component. While still large, this component is scalable, because for one thing high $q^2$ is the preferred kinematic region for lattice-QCD calculations.

We find that, over the long haul, the total error for $\Bsmmy$ at high $q^2$ is dominated by the experimental component. Absent an analysis optimized for the $\Bsmmy$ search, we estimate this error from the sheer statistical component. We then infer the $\Bsmmy$ sensitivity as the distance of the SM prediction vs. the prediction within the aforementioned NP benchmarks, in units of the total error determined as described. In the example of the real-$\delta C_{9,10}$ scenario, we find that the pull as a function of the acquired data reaches the $2\sigma$ level at the border of the $1\sigma$ region for the Wilson-coefficient shift.

In case such sensitivity may look underwhelming, we make the following final remarks. First, the above sensitivity is to be compared with other {\em single}-observable sensitivities that one can expect from e.g. BRs and angular analyses. For observables at low $q^2$, sensitivities that are quoted in the literature tacitly rely on a breakthrough in the understanding of long-distance effects. As already emphasized, this issue is absent at {\em high} $q^2$ in the case of \Bsmmy. Yet, we are not aware of other detailed studies of high-$q^2$ exclusive observables to compare our study against. An interesting (semi-)inclusive example is the recent Ref.~\cite{Isidori:2023unk}.

Second, in all likelihood any NP in semi-leptonic $B$ decays will first be established {\em collectively}, i.e. through many modes showing a coherent trend---and a persistent one with increasing statistics. Only later will such trend be consolidated by single observables getting to the canonical $5\sigma$ departures required.
If this is the case, then it is of the highest importance to find as many measurables as possible that allow to confirm the trend---e.g. experimental branching ratios below the theory prediction---in observables devoid of the long-distance issues that at present plague low $q^2$. This study goes in this direction.

\section*{Acknowledgments} We warmly thank Martino Borsato for useful feedback, and Alexandre Carvunis for help on various aspects of {\tt flavio}. We are grateful to Francesco Dettori for several discussions and for input related to Sec. \ref{sec:Bsmmy_exp}. This work is supported by ANR under contract n. 202650 (ANR-19-CE31-0016, GammaRare).

\appendix

\section*{Appendix: Global fits in SMEFT}\label{app:SMEFT}

\begin{figure}[th!]
  \begin{center}
    \includegraphics[width=0.43\linewidth]{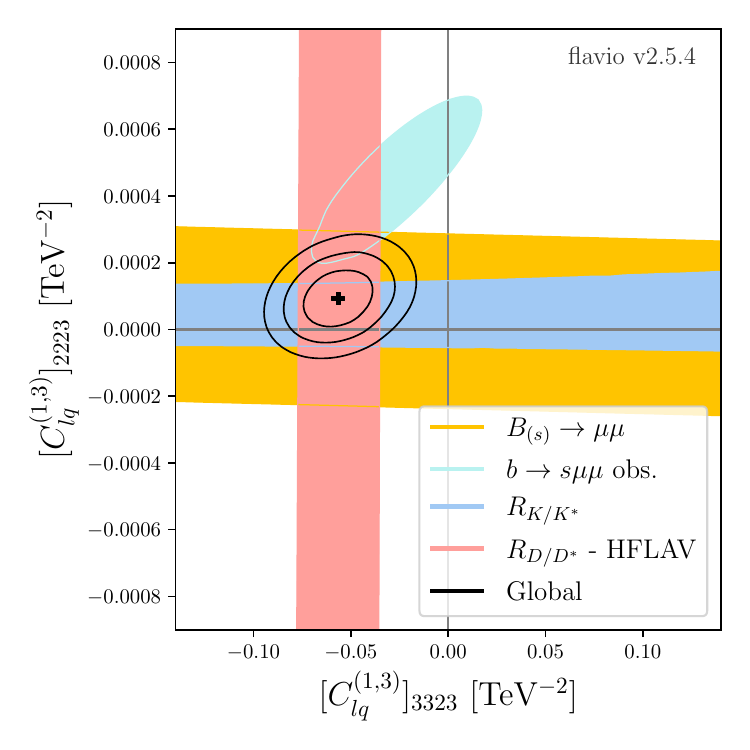} \hfill
    \includegraphics[width=0.43\linewidth]{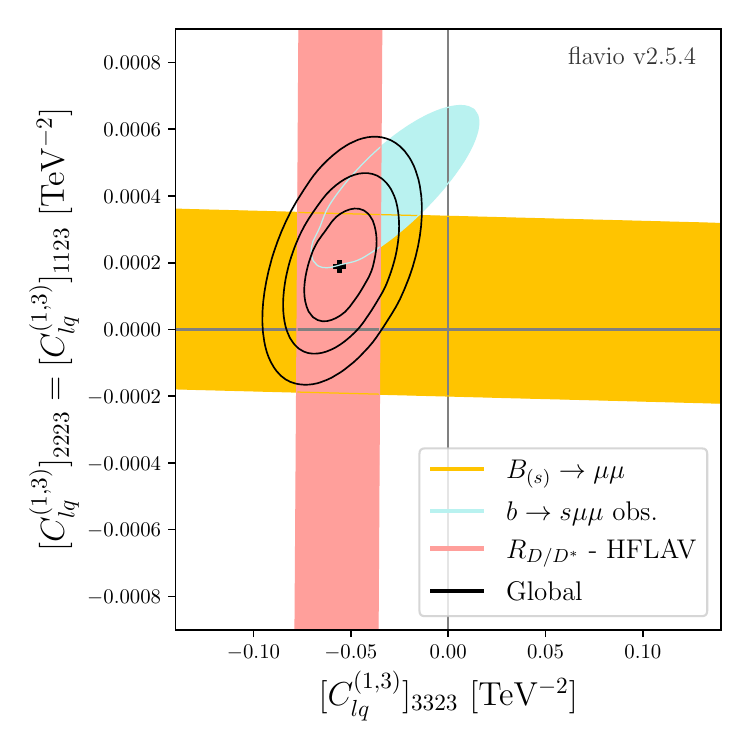}
    \caption{Counterparts of the top-left and top-right panels of \reffig{fig:fit-RDDst}, respectively, in the plane of the underlying SMEFT Wilson coefficients.
    }
  \label{fig:SMEFT-fits}
    \end{center}
\end{figure}

In Fig.\,\ref{fig:SMEFT-fits}, we collect the counterparts of the top-left and top-right panels of \reffig{fig:fit-RDDst}, but in the plane of the underlying SMEFT Wilson coefficients. For the sake of self-consistency, we do not use here the abbreviations introduced in \refsec{sec:SMEFT} and instead simply stick to the Warsaw-basis notation \cite{Grzadkowski:2010es}.

\bibliography{bibliography}
\bibliographystyle{JHEP}

\end{document}